\documentclass[twocolumn]{aastex631}

\usepackage{amsmath}

\shorttitle{Detecting IMBHs Using Quasar Microlensing}
\shortauthors{WU \& HO}

\begin{document}

\title{Detecting Intermediate-mass Black Holes Using Quasar Microlensing}

\correspondingauthor{Zihao Wu}
\email{zihao.wu@cfa.harvard.edu}

\author[0000-0002-8876-5248]{Zihao Wu}
\affiliation{Center for Astrophysics {$\vert$} Harvard \& Smithsonian, 60 Garden Street, Cambridge MA 02138, USA}
\affiliation{Kavli Institute for Astronomy and Astrophysics, Peking University, Beijing 100871, China}
\affiliation{Department of Astronomy, School of Physics, Peking University, Beijing 100871, China}

\author[0000-0001-6947-5846]{Luis C. Ho}
\affiliation{Kavli Institute for Astronomy and Astrophysics, Peking University, Beijing 100871, China}
\affiliation{Department of Astronomy, School of Physics, Peking University, Beijing 100871, China}

\begin{abstract}

Recent studies suggest that numerous intermediate-mass black holes (IMBHs) may wander undetected across the Universe, emitting little radiation. These IMBHs largely preserve their birth masses, offering critical insights into the formation of heavy black hole seeds and the dynamical processes driving their evolution. We propose that such IMBHs could produce detectable microlensing effects on quasars. Their Einstein radii, comparable to the scale of quasar broad-line regions, magnify radiation from the accretion disk and broad emission lines, making these quasars outliers in flux scaling relations. Meanwhile, the microlensing causes long-term, quasi-linear variability that is distinguishable from the stochastic variability of quasars through its coherent multi-wavelength behavior. We develop a matched-filtering technique that effectively separates the long-term lensing signal from the intrinsic quasar variability, with sensitivity tripling each time the observational time span doubles. Moreover, as IMBHs are often surrounded by dense star clusters, their combined gravitational field produces substantial extended, concentric caustics. These caustics induce significant variability in optical, ultraviolet, and X-ray bands over decade timescales, alongside hour-to-day-scale flux fluctuations in broad emission lines. We predict a substantial number of detectable events in the upcoming surveys by the Vera C. Rubin Observatory, considering recent IMBH mass density estimates. Even in the absence of positive detections, searches for these microlensing signals will place meaningful constraints on the cosmological mass density of IMBHs, advancing our understanding of their role in cosmic evolution.
\end{abstract}

\keywords{gravitational lensing: micro -- galaxies: active -- galaxies: nuclei -- quasars}

\section{Introduction} \label{sec:intro}

Intermediate-mass black holes (IMBHs), commonly defined in the mass range $10^2-10^6\,M_\odot$, bridge the gap between stellar-mass black holes and supermassive black holes. They are viewed as seeds for supermassive black holes, making their abundance crucial to understanding the growth history of quasars in the early Universe \citep{Fan2023, Pacucci2023ApJ}, a subject of increasing interest in light of the high abundance of accreting black holes recently emerging from observations with the James Webb Space Telescope (e.g., \citealt{harikane_2023, greene_2024, matthee_2024}). However, the cosmological abundance of IMBHs remains poorly understood.

Numerous efforts have been devoted to finding IMBHs as accreting nuclei in low-mass, late-type galaxies. To date, substantial samples have been accumulated through systematic surveys in the optical \citep{greene_active_2004, greene_2007, dong_2012, liu_2025} and X-rays (\citealt{desroches_2009, gallo_2010, Miller2015ApJ, she_2017, bi_2020}). Deep, high-resolution imaging in the radio by \cite{reines_2020} has also yielded promising candidates. \cite{greene_intermediate-mass_2020} conclude that at least 50\% of galaxies in the local Universe with stellar mass below $10^{10}\,M_\odot$ harbor black holes in the IMBH regime. 

However, accreting IMBHs in galaxy centers constitute only a small portion of the overall IMBH population, whereas the vast majority may wander in galaxy halos due to insufficient dynamical friction \citep{ricarte_origins_2021}, galaxy minor mergers \citep{greene_intermediate-mass_2020}, and ejection via three-body interactions in galactic nuclei \citep{Fragione2020MNRAS}. These wandering IMBHs are predicted to be highly abundant in cosmological hydrodynamical simulations \citep{ricarte_origins_2021, weller_dynamics_2022, di_matteo_vast_2022}, although significant discrepancies remain between different simulations \citep{Donkelaar2024arXiv}.

Detecting wandering IMBHs remains challenging due to their faintness. Recent searches through the dynamics of globular clusters \citep{aros_dynamical_2020} and hypercompact star clusters \citep{greene_search_2021} have not yielded concrete evidence. Advancements in observational techniques may allow future detection through X-ray, millimeter-wave, and gravitational wave emissions from IMBHs \citep{cutler_angular_1998, fragione_tidal_2018, greene_intermediate-mass_2020, Guo2020ApJ}. However, these methods are subject to significant selection biases related to IMBH accretion physics and dynamical environments.

Gravitational microlensing offers an alternative method to detect wandering IMBHs based solely on their masses. Historically, microlensing provided critical evidence that ruled out massive compact halo objects as the primary component of dark matter \citep{paczynski_gravitational_1986, alcock_macho_2000}. Moreover, microlensing has become a cornerstone in exoplanet discovery \citep{zhu_exoplanet_2021} and led to the discovery of the first isolated stellar-mass black hole \citep{sahu_isolated_2022}.

The term quasar microlensing traditionally refers to the brightness perturbation of strongly lensed quasars caused by microlensing from stars in the lens galaxies \citep{kochanek_quantitative_2004, Schmidt_Wambsganss_2010}. Stellar microlensing may form complex caustics, leading to quasar variability on time scales of months \citep{Schmidt_Wambsganss_2010}. This effect has been utilized to study the size and brightness profiles of the inner structures of quasars \citep[e.g.,][]{morgan_2010_apj}.
 
More broadly, quasar microlensing includes any microlensing effects from foreground objects, not limited to strong lensing scenarios. In this context, quasar microlensing has been proposed as a method to detect foreground dark matter in the stellar-mass range, which produces microlensing signal on time scales of months to years \citep{dalcanton_observational_1994, gould_microlensing_1995, mediavilla_limits_2017, awad_probing_2023, perkins_disentangling_2024}. Additionally, \cite{hawkins_gravitational_1993} suggested that the year-scale variability of quasars is primarily caused by microlensing from objects with masses around $0.05\,M_\odot$, although this interpretation remains controversial \citep{baganoff_gravitational_1995, hawkins_quasar_1997}.

Here we demonstrate that quasars serve as excellent background sources for detecting microlensing by IMBHs. They are ideal isolated point sources for IMBH microlensing, and their vast distances allow probing a large volume, thereby enhancing the event rates. Quasar microlensing by IMBHs has a substantially larger Einstein radius compared to those produced by stellar objects, resulting in distinct observational characteristics due to the reduced prominence of finite source effects. Furthermore, IMBHs are usually surrounded by stellar clusters (e.g., \citealt{Filippenko2003, greene_intermediate-mass_2020}), producing caustic curves of considerable size and unique patterns.

Many tentative microlensing signals of IMBHs have been reported. \cite{krol_possible_2023} found that the peculiar achromatic light curve of an active galaxy is consistent with binary microlensing involving an IMBH. \cite{paynter_evidence_2021} reported a time delay in a gamma-ray burst that is possibly caused by IMBH lensing. Other studies have proposed methods to detect IMBHs through their perturbation to the Einstein radius of strongly lensed quasars \citep{inoue_direct_2013}, astrometric microlensing effects on stars in host globular clusters \citep{Kains2016MNRAS, Kains2018ApJ}, and gravitational wave lensing \citep{gais_inferring_2022, meena_gravitational_2023}. 

Previous IMBH microlensing events have mostly been discovered fortuitously. With the advent of the Large Synoptic Survey Telescope (LSST; \citealt{ivezic_lsst_2019}), we are entering a new era with unprecedented capabilities in wide-area time-domain surveys, offering the potential to detect many more such events in the near future. It becomes possible to constrain the cosmological abundance of IMBHs through systematic search of IMBH microlensing signals.

We present the basic observational properties in Section~\ref{sec:obs_prop} and approaches to distinguishing the microlensing signals from contaminants in Section~\ref{sec:multiband}. We estimate the number of events under conditions anticipated for upcoming observational facilities in Section~\ref{sec:event_number}. We introduce a new technique for detecting the lensing signal and evaluate the detection limit in Section~\ref{sec:detectability}. Results are summarized in Section~\ref{sec:summary}. This study adopts a $\Lambda$CDM cosmology with $H_0 = 69\,\mathrm{ km\, s^{-1} \,Mpc^{-1}}$, $\Omega_m = 0.29$, and $\Omega_\Lambda = 0.71$ according to the Nine-year Wilkinson Microwave Anisotropy Probe Observations \citep{hinshaw_nine-year_2013}. Throughout we interchangeably use the terms ``quasar'' and ``active galactic nucleus'' (AGN).

\section{Observational Properties}
{
\label{sec:obs_prop}
This section presents statistical properties of IMBH microlensing. We predict the statistical distribution of IMBH Einstein radii, microlensing variability amplitudes, and the average magnification rate in observed events. Additionally, we analyze the characteristics of the observed segments of microlensing light curves. Finally, we discuss caustic effects induced by possible stellar clusters surrounding the IMBH.

\subsection{Einstein Radius}
\label{sec:einstein radii}
The Einstein radius, $\theta_\mathrm{E}$, determines the angular scale of gravitational lensing. For a lensing IMBH of mass $M$ at redshift $z_l$ and a background quasar at $z_s$, it is expressed as

\begin{equation}
\theta_\mathrm{E} = \sqrt{\frac{4GM(1+z_l)}{c^2} \frac{D_{ls}}{D_l D_s}},
\label{eq:einstein_radii}
\end{equation}

\noindent
where $D_l$ and $D_s$ are the comoving distances from the IMBH and the quasar to Earth, respectively, and $D_{ls}=D_s-D_l$ is the comoving distance between the IMBH and the quasar \citep{schneider_gravitational_2006}. 

Using Monte Carlo simulations assuming a constant comoving density of IMBHs across redshifts \citep{ricarte_origins_2021, di_matteo_vast_2022}, we compute the probability distributions of $\theta_\mathrm{E}$ and its projected physical size $r_\mathrm{E} = \theta_\mathrm{E} D_s/(1+z_s)$ on the quasar plane for IMBHs with mass $M=10^3\,M_\odot$. As Figure~\ref{fig:thetaE} shows, for $z_s \gtrsim 0.5$, $\theta_\mathrm{E}$ stabilizes near 0.05 mas, corresponding to $r_\mathrm{E} \approx 0.4\ \mathrm{pc}$. 

For comparison, the sizes of quasar accretion disks, as measured from lensing, are usually $10^{-4}$ to $10^{-2}$\,pc \citep{morgan_2010_apj, Vernardos_2024_SSRv}, and the sizes of the broad-line region (BLR) in quasars, determined from reverberation mapping, range from $ 10^{-3}$ to $10^{-1}$\,pc \citep{kaspi_relationship_2005, bentz_2013, du_2016, li_2021}. Although the exact values vary with the black hole mass and quasar luminosity, our estimates suggest that the Einstein radius is significantly larger than the accretion disk and slightly exceeds the size of most BLRs. The implications of this will be further explored in Section~\ref{sec:multiband}.

Current optical telescopes lack the resolution to detect such small angular scales. However, radio interferometers like the Event Horizon Telescope with $0.02\ \mathrm{mas}$ resolution \citep{EHT2019} or optical/near-infrared instruments like GRAVITY with sub-mas resolution \citep{GRAVITY2017A&A, Eisenhauer_2023_ARAA} could resolve these signals in the future. While flux sensitivity remains a challenge for high-redshift targets, the long timescales of quasar microlensing allow for future developments of observational techniques to examine the effect.

\begin{figure}
\includegraphics[width=0.98\linewidth]{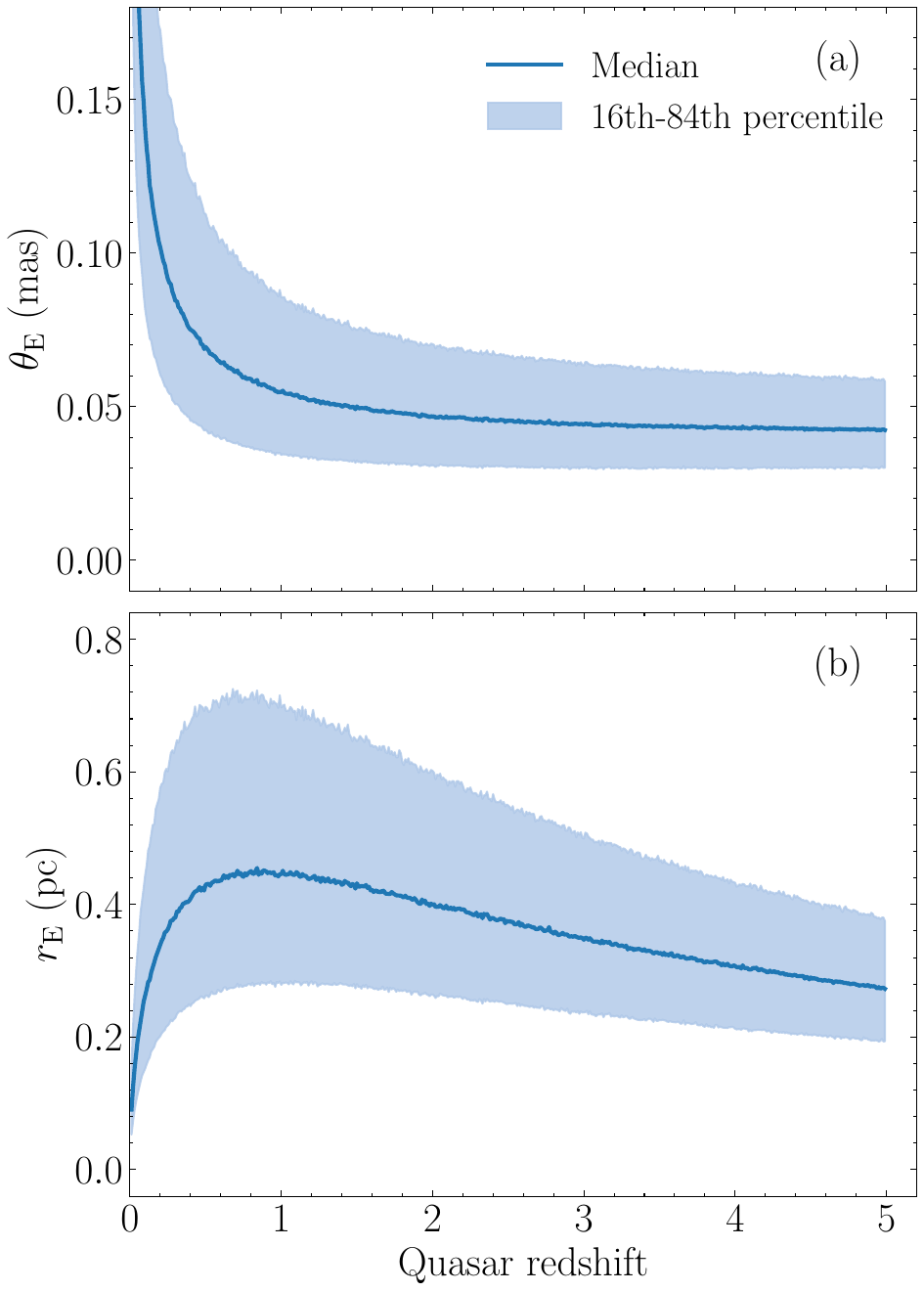}
\caption{Redshift variation of (a) the Einstein radius $\theta_\mathrm{E}$ and (b) its projected size $r_\mathrm{E}$ in the quasar plane. For a quasar at a given redshift, we simulate the probability distributions of $\theta_\mathrm{E}$ and $r_\mathrm{E}$ assuming a constant comoving number density of IMBHs. We show the median value of the distribution with a solid line and the 16th--84th percentile ($1\,\sigma$) range with the shaded region. We consider an IMBH with mass $M = 10^3\, M_\odot$; the results for other masses scale by a factor of $(M/10^3\,M_\odot)^{1/2}$.}
\label{fig:thetaE}
\end{figure}

\subsection{Magnification Amplitude and Variability}
\label{sec:magnitude_variation}

Microlensing affects quasar brightness via magnification and variability. The magnification rate is 

\begin{equation}
A = \frac{2+u^2}{u\sqrt{u^2+4}},
\label{eq:magnification}
\end{equation}

\noindent
where $u$ is the angular separation between the quasar and the lens normalized by the Einstein radius. We denote $m_\mathrm{lens}=-2.5\log A$ as the microlensing magnification in magnitude units and define the instantaneous microlensing variability as

\begin{equation}
\begin{aligned}
\mathcal{A} \equiv \left|\frac{1}{A}\frac{dA}{dt}\right| = \frac{8}{u(u^2+4) (u^2+2)} \left|\frac{\mathrm{d}u}{\mathrm{d}t} \right| .
\end{aligned}
\label{eq:variability}
\end{equation}

\noindent
As $m_\mathrm{lens}=-1.09\ln A\approx-\ln A$, we further obtain $\mathcal{A}\approx \left|{\mathrm{d}m_\mathrm{lens}}/{\mathrm{d} t}\right|$. The dependence on $\mathrm{d}u/\mathrm{d}t$ indicates that the microlensing variability is proportional to the relative velocity between quasars and IMBHs in Einstein radius units.

\begin{figure*}
\centering
\includegraphics[width=0.98\linewidth]{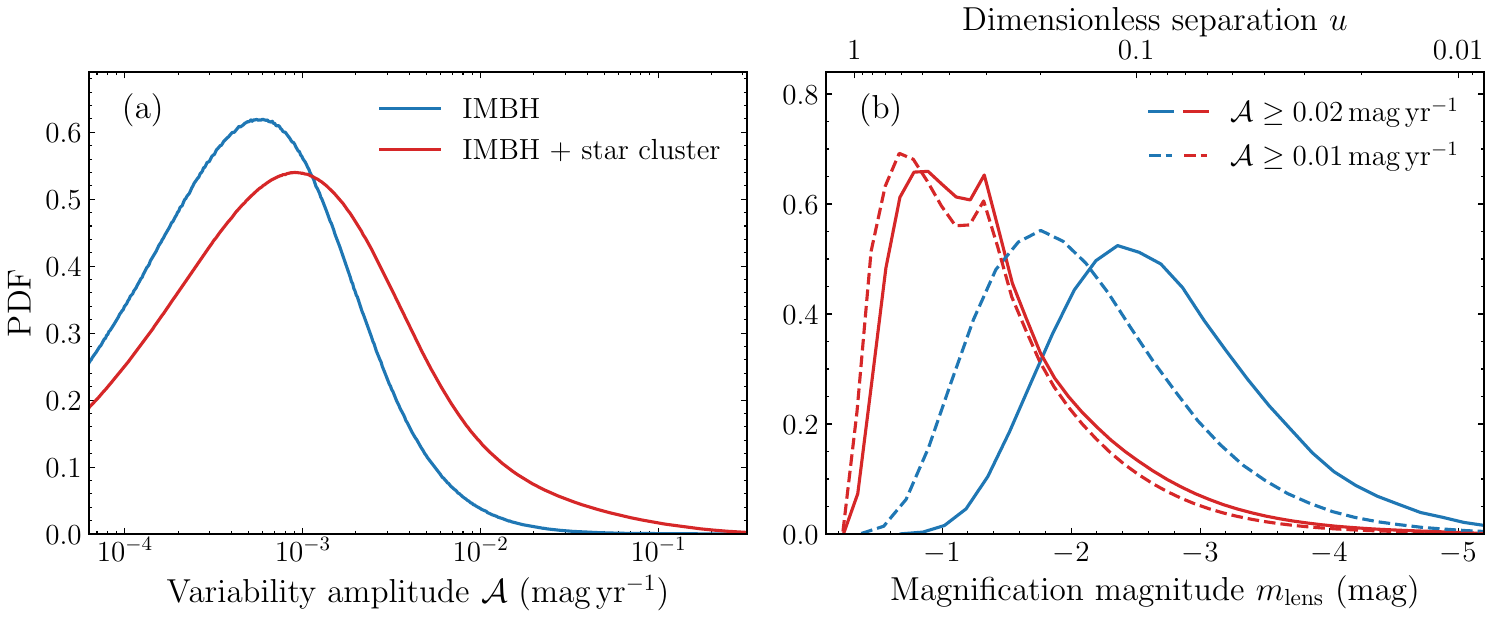}
\caption{(a) Probability density function (PDF) of microlensing variability amplitude $\mathcal{A}$ for an isolated IMBH (blue curve; Section~\ref{sec:magnitude_variation}) and an IMBH with a companion star cluster (red curve; Section~\ref{sec:caustics}). (b) Conditional PDF of lensing magnification magnitude $m_\mathrm{lens}$ for events with variability amplitude exceeding $\mathcal{A} \geq 0.02$\,mag\,yr$^{-1}$ (solid lines) and $\mathcal{A} \geq 0.01$\,mag\,yr$^{-1}$ (dashed lines). Blue and red curves correspond to cases of an isolated IMBH and an IMBH with a star cluster, respectively. The twin axis on the top is the dimensionless separation $u$ corresponding to the magnification magnitude in single lens, with their relation given by Equation~\ref{eq:magnification}. Realistic detection thresholds for $\mathcal{A}$ are discussed in Section~\ref{sec:detectability}.}
\label{fig:variability_and_magnification}
\end{figure*}

The relative velocity comes from several astrophysical components. First, host galaxies of quasars and IMBHs have peculiar velocities with respect to their local Hubble flow. \cite{hawkins_2df_2003} show that the peculiar velocities follow an exponential distribution with a typical dispersion of $600\,\mathrm{km\,s^{-1}}$. Second, IMBHs wandering in galaxies have a virial velocity of $100-1000\,\mathrm{km\,s^{-1}}$ relative to the galactic center depending on the halo mass \citep{Brimioulle_2013_MNRAS}. Third, the formation processes of IMBHs may impart additional peculiar velocities. Simulations suggest that wandering IMBHs originating from collisions of dwarf galaxies with the Milky Way acquire velocities of $\sim 100\,\mathrm{km\,s^{-1}}$ \citep{weller_dynamics_2022}, while IMBHs formed in globular clusters have a high probability of being ejected with velocities of a few $100\,\mathrm{km\,s^{-1}}$ \citep{holleybockelmann_gravitational_2008}. In rare cases, gravitational recoil can accelerate an IMBH up to $4000\,\mathrm{km\,s^{-1}}$ \citep{Campanelli_PRL_2007}. Finally, the solar system's peculiar motion of $369\,\mathrm{km\,s^{-1}}$ contributes to the relative motion via parallax effects \citep{hinshaw_five-year_2009}.

\begin{figure}[b]
\centering
\includegraphics[width=0.98\linewidth]{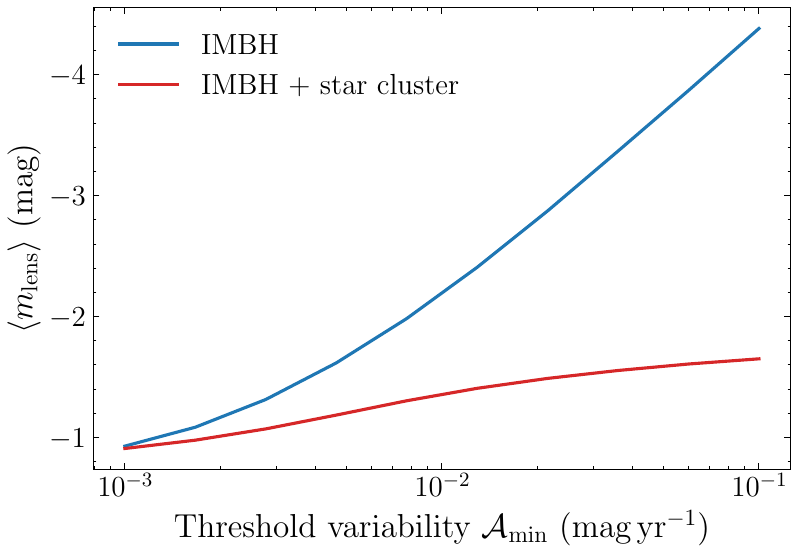}
\caption{The mean lensing magnification $\langle m_\mathrm{lens}\rangle$ for events exceeding thresholds of microlensing variability amplitude $\mathcal{A}_\mathrm{min}$, for an isolated IMBH (blue curve; Section~\ref{sec:magnitude_variation}) and an IMBH with a star cluster (red curve; Section~\ref{sec:caustics}).}
\label{fig:mean_magnification}
\end{figure}

As the cosmological peculiar velocities contribute most to the velocity budget, we simplify the velocity distribution as an exponential with typical values of $600\,\mathrm{km\,s^{-1}}$ and $700\,\mathrm{km\,s^{-1}}$ for quasars and IMBHs, respectively. We adopt slightly higher velocities for IMBHs to account for their prevalent peculiar motions relative to their host galaxies. We find a typical relative velocity in Einstein radius units of $\mathrm{d}u/\mathrm{d}t=10^{-3}$\,yr$^{-1}$ for an IMBH at $z_l=1$ and a quasar at $z_s=2$, considering random velocity orientation and cosmological time dilation.

Using Monte Carlo simulations, we predict the probability density function (PDF) of the microlensing variability amplitude $\mathcal{A}$. We account for random distribution of velocity amplitudes and directions, with a mean of $\mathrm{d}u/\mathrm{d}t=10^{-3}$\,yr$^{-1}$. We only consider events with $u\leq 1$, with $u$ randomly sampled in the plane. As shown by the blue curve in Figure~\ref{fig:variability_and_magnification}a, $\mathcal{A}$ is concentrated near $10^{-3}$\,mag\,yr$^{-1}$, with only a small fraction exceeding $10^{-2}$\,mag\,yr$^{-1}$. However, we will demonstrate in Section~\ref{sec:detectability} that only those with $\mathcal{A}\gtrsim10^{-2}$\,mag\,yr$^{-1}$ are practically observable through variability. Therefore, the observed PDF of $\mathcal{A}$ is only the right-side tail of the overall PDF in Figure~\ref{fig:variability_and_magnification}a.

\begin{figure*}
\centering
\includegraphics[width=0.98\textwidth]{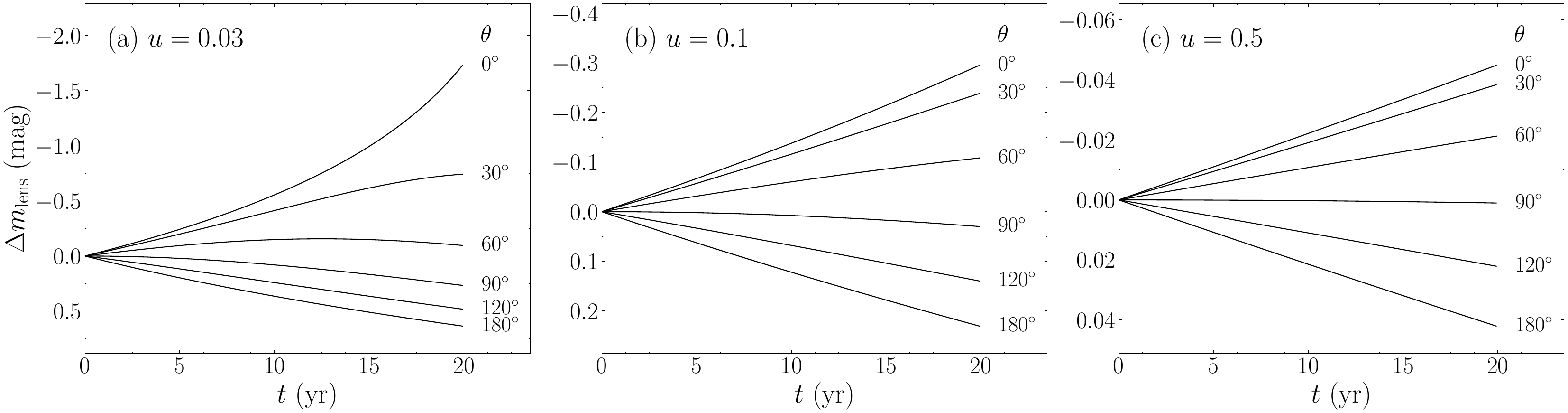}
\caption{Illustration of the observable part of microlensing light curves over a time span of 20 years. The variation of microlensing magnitude induced by the relative motion between the quasar and IMBH is denoted by $\Delta m_\mathrm{lens}$, and $t$ indicates the time since the start of the observational monitoring. The upper left corner of each subplot shows the dimensionless angular separation $u$ at time $t=0$. The angle $\theta$ represents the angle between the relative velocity and the line connecting the IMBH and the quasar, with the head-on direction corresponding to $\theta=0^\circ$. This example assumes a quasar at redshift $z_s=2$ lensed by an IMBH with a mass $M=10^3\,M_\odot$ located at redshift $z_l=1$.}
\label{fig:lens_lc}
\end{figure*}

We further analyze the conditional PDF of the lensing magnification magnitude $m_\mathrm{lens}$ for events above thresholds of $\mathcal{A} = 0.02\,$mag\,yr$^{-1}$ and $0.01\,$mag\,yr$^{-1}$. Figure~\ref{fig:variability_and_magnification}b shows that those events with high variability are accompanied by strong magnification and small $u$. The average magnification rate reaches 3\,mag for a threshold of $\mathcal{A} \geq 0.02\,\rm mag\,yr^{-1}$ and 2\,mag for $\mathcal{A} \geq 0.01\,\rm mag\,yr^{-1}$. Their typical $u$ is merely $\sim0.1$. The dependence of average magnification rate on variability thresholds (Figure~\ref{fig:mean_magnification}) suggests that microlensing events detected through high variability will preferentially trace high-magnification events.

We generalize our results to other configurations of redshifts and velocities using the following scaling relation: when expressed in units of the Einstein radius, the variability $\mathcal{A} = \mathcal{A}(u, \mathrm{d}u/\mathrm{d}t)$ is independent of any physical scale, as a consequence of which events with different relative velocities $v$ and Einstein radii $r_\mathrm{E}$ have the same variability $\mathcal{A}$ as long as the ratio $v/r_\mathrm{E}$ remains constant. Specifically, since $\mathcal{A} \propto \mathrm{d}u/\mathrm{d}t$, we have the scaling relation

\begin{equation} \mathcal{A} \propto \frac{v}{r_\mathrm{E}(1+z)} \propto \frac{1}{\sqrt{M}}. \label{eq:scale} \end{equation}
\noindent

\noindent
This relation implies that the PDF of $\mathcal{A}$ in configurations of different relative velocities, Einstein radii, redshifts, and IMBH masses can be obtained by rescaling the $x$-axis of the distribution in Figure~\ref{fig:variability_and_magnification}a.

\subsection{Microlensing Light Curve}

In the common context of microlensing, one observes a full light curve showing a characteristic rise and fall. For quasar microlensing by IMBHs, however, only a short segment of the light curve is observed in any realistic observations. This segment appears as a nearly linear trend in the quasar light curve given the long timescale. We demonstrate this effect in this section and discuss when higher order curvature terms become relevant.

Figure~\ref{fig:lens_lc} shows the microlensing light curves of a representative case: a quasar at $z_s=2$ lensed by an IMBH of mass $M=10^3\,M_\odot$ at $z_l=1$, with a typical relative angular velocity $\mu_\mathrm{rel} = 0.05\,\mathrm{\mu as\,yr^{-1}}$ and a range of orientations $\theta$, with $\theta=0^\circ$ when the quasar and IMBH move toward each other. As expected, over the first 10 years the light curve is essentially linear. Even after 20 years, while slight curvature appears in some cases, the linear approximation remains remarkably accurate.

To quantify this, we expand the microlensing magnitude variation in a Taylor series. In the limit that $u\rightarrow 0$ and $\Delta u/u \rightarrow 0$,

\begin{equation}
\begin{aligned}
  m_\mathrm{lens} &\approx -\ln ||\vec{u}+\Delta \vec{u}|| = -\frac{1}{2}\ln(u^2+\Delta u^2 - 2u\Delta u \cos \theta)\\
   &=-\ln u + \frac{\Delta u}{u} \cos \theta + \frac{1}{2} \left(\frac{\Delta u}{u}\right)^2 \cos 2 \theta +\mathcal{O}(3) ,
\end{aligned}
\end{equation}

\noindent
where $\Delta \vec{u}$ is the displacement of $u$ in the relative velocity direction. The ratio of the second-order term to the linear term is

\begin{equation}
R = \frac{{\Delta u \cos2\theta}}{2u\cos \theta} \approx \frac{\cos2\theta}{2\cos^2\theta}\Delta {m_\mathrm{lens}},
\end{equation}
\noindent

\noindent
where we have used $\Delta m_\mathrm{lens} \approx (\Delta u/u)\cos\theta $ as the first-order approximation. Therefore, only when $\Delta m_\mathrm{lens}>1\,$mag or when the relative motion is tangential, do higher order terms become significant. However, the first condition usually takes a long time to achieve due to the typically small variability, and the latter case is practically undetectable, as the first order term of $m_\mathrm{lens}$ vanishes, resulting in extremely little variability.

Therefore, for the majority of detectable events, the microlensing variability presents as a linear light curve. The higher order curvature is only observable after long observation time span to accumulate $\Delta m_\mathrm{lens}$. While linear light curves are invariant under time translation, the presence of higher order terms breaks this symmetry, thereby enabling the detection of time delays between light curves. This further helps to distinguish microlensing signals from quasar intrinsic variability, which usually has wavelength-dependent time delays (Section~\ref{sec: simutaneous}). 

\begin{figure*}
\centering
\includegraphics[width=0.98\textwidth]{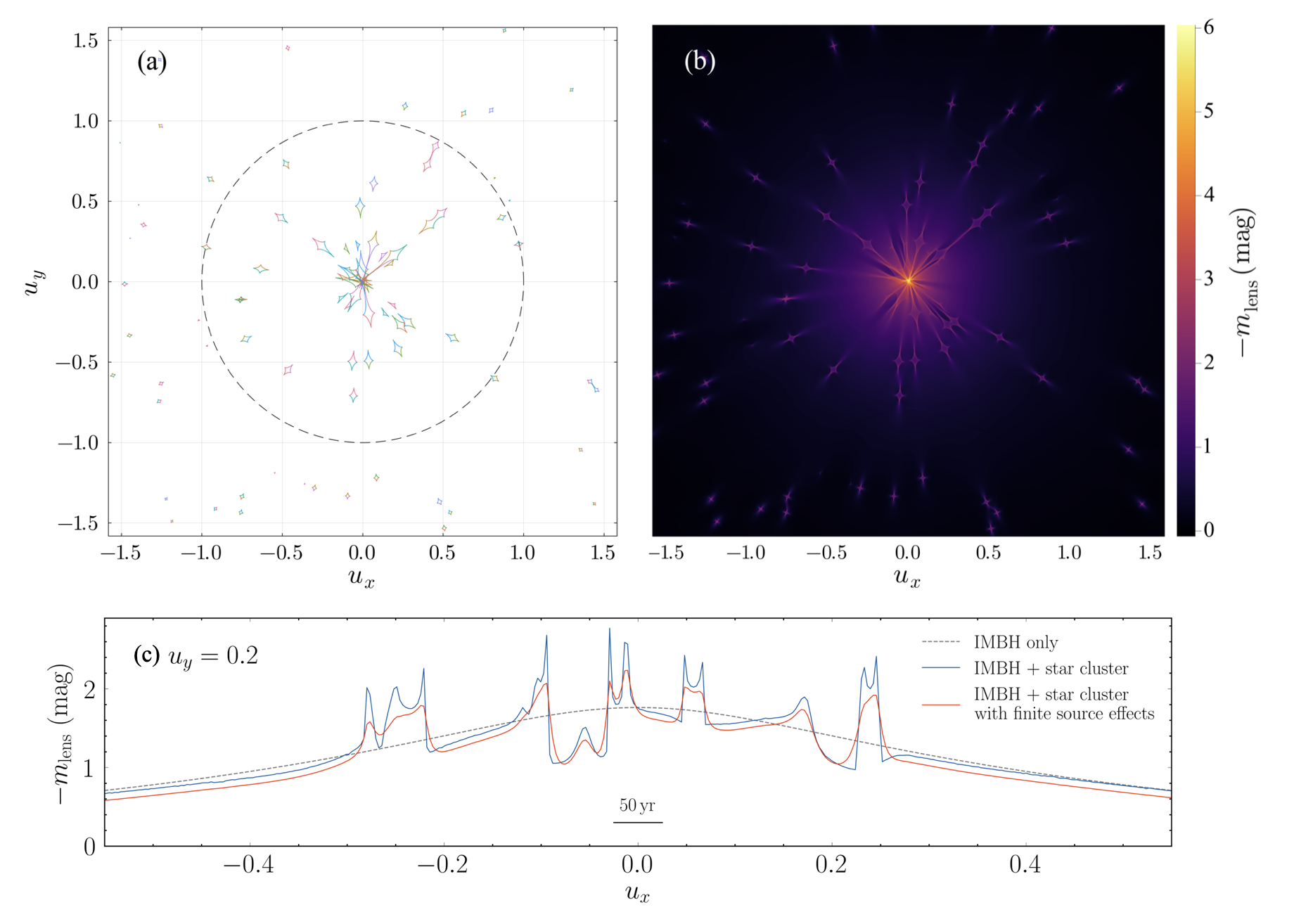}
\caption{Effects of a star cluster surrounding an IMBH located at the origin, with the stars randomly distributed near the IMBH. Panel (a) shows the microlensing caustics and the Einstein ring of the IMBH (dashed circle); the angular distances $u_x$ and $u_y$ are given relative to the IMBH, normalized by its Einstein radius. The magnification map in panel (b) is colored to indicate the lensing magnification magnitude $m_\mathrm{lens}$. Panel (c) presents the variation in magnification magnitude along $u_x$ at $u_y = 0.2$ with (red solid line) and without (blue solid line) finite source effects of the quasar accretion disk, in comparison to the smoother magnification curve (grey dashed line) from the IMBH alone. This curve also represents the light curves of quasars moving along the $u_x$ direction, with the scale bar representing changes in $u_x$ over 50 years, given typical relative velocities between IMBHs and quasars, as discussed in the text.}
\label{fig:caustics}
\end{figure*}

\subsection{Caustics by IMBHs and Star Clusters}
\label{sec:caustics}

IMBHs often reside within dense stellar clusters \citep{gebhardt2005,fragione2018, greene_intermediate-mass_2020, Lena2020MNRAS}, which can produce intricate caustic structures in addition to the main Einstein ring of the IMBH. These caustics introduce additional variability as a quasar crosses them. We simulate the caustic features with the package {\tt microlensing.jl} \citep{Pirogov2020}, where we consider a star cluster around an IMBH of mass $10^3\, M_\odot$ at $z_l =1$, with an Einstein radius of 0.05\,mas (Figure~\ref{fig:thetaE}). Stars are modeled with masses of 1\,$M_\odot$ and a uniform surface density of 30\,$\mathrm{pc^{-2}}$, consistent with dynamical simulations that suggest a flat two-dimensional density profile near the IMBH, ranging from 1 to 200\,$\mathrm{pc^{-2}}$ 
\citep{Baumgardt_2004ApJ1,Baumgardt_2004ApJ2}.

The star cluster produces radially oriented, elongated caustic curves concentrated near the IMBH (Figure~\ref{fig:caustics}a). Moreover, the IMBH dominates the overall magnification map, while the stellar caustics introduce localized, highly magnified regions (Figure~\ref{fig:caustics}b). A cut through this map (Figure~\ref{fig:caustics}c) shows that these caustics produce flare-like features superimposed on the longer, smoother variability induced by the IMBH; these flares can reach $\gtrsim 1$\,mag over timescales of years. Between flares, the light curve shows characteristic ``U-shaped" depressions, similar to binary microlensing \citep{alcock_binary_2000}, where the total magnification briefly drops but remains above the IMBH-only magnification. 

We further consider the finite source effects of quasar accretion disks. We convolve the magnification map with the surface brightness profile of a standard, optically thick, geometrically thin accretion disk \citep[$T\propto r^{-3/4}$;][]{shakura_1973} in the face-on direction. The surface brightness profile is given by the Planck function \citep{Poindexter2008ApJ}:

\begin{equation}
\Sigma (r) = \frac{2h\nu^3}{c^2}\frac{1}{e^{(h\nu/k_BT)}-1}\propto \frac{1}{e^{a(r/r_\mathrm{half})^{3/4}}-1},
\end{equation}

\noindent
where $a$ is a normalization factor that ensures that half of the total flux lies within $r_\mathrm{half}$. We adopt $r_\mathrm{half}=5 \times 10^{15}$\,cm as measured in the $B$ band \citep{Poindexter2008ApJ} and a typical $r_\mathrm{E}=0.4$\,pc (Figure~\ref{fig:thetaE}). 

Finite source effects smooth out the sharp brightness peaks and reduce the amplitude of the flares (Figure~\ref{fig:caustics}c, red curve). As the effective radius of the accretion disk increases with wavelength, the finite source effect is stronger in the redder bands. Therefore, the amplitude of lensing variability and magnification during caustic crossing, whose timescale lasts decades, should decrease progressively from the X-rays to the ultraviolet and optical bands, while the timescale is expected to increase, although the time interval between consecutive caustic crossing events will be similar across wavelengths. 

Finite source effects on the BLR are more complicated because the line-emitting gas consists of many small, fast‐moving clouds, whose individual sizes are much smaller than the caustic curves. This means that the finite source effect for each cloud is negligible, rendering significant magnification of individual clouds possible. The caustic crossings of clouds have much shorter timescales: for a cloud with a size of $10^{12}\,\mathrm{cm}$ \citep{Laor2006ApJ} moving at a speed of $\sim10^3\,\mathrm{km\,s}^{-1}$ \citep{li_bayesian_2013}, the caustic crossing timescale is merely $\sim 3$ hr. Cosmological time dilation and velocity projection would increase this moderately, such that the timescale may range from hours to a day.

With a total spatial extent of $10^{-3}$–$10^{-1}$\,pc \citep{du_2016, li_2021}, the BLR spans $\sim$1\%--10\% of the size of the typical Einstein radius. Meanwhile, the caustic size is $\sim \sqrt{q}\,r_\mathrm{E}\approx 0.03r_\mathrm{E}$, where $q=0.001$ is the mass ratio between the star and the IMBH \citep{Han2006ApJ}. The large size of the BLR suggests that several caustic curves may cross it simultaneously, especially when the quasar is close to the lens centroid where the caustics are dense (Figure~\ref{fig:caustics}a). With a covering fraction of at least 10\% in quasars \citep{Goad2012}, the BLR is sufficiently densely distributed that many clouds may be on the caustic curves and produce flares simultaneously. Although the flare from each cloud is diluted by the entire BLR, the collective effect of numerous caustic crossings may be observed statistically. Given the stochastic motion of the clouds, each caustic crossing event should occur randomly in time. These flares, when added together, behave as a shot noise process \citep{Thorne2017mcp} and induce flux fluctuations with a power spectrum peaking around the timescale of each single caustic crossing event. We can thus identify this effect from BLR flux variations on hour-to-day timescales. The flux contribution from clouds experiencing caustic crossing may be sufficiently pronounced to induce changes in the line profile of the broad emission lines. If the cloud motions are predominantly Keplerian with aligned angular momentum, the similar velocity direction of the magnified clouds may even produce an overall shift to the line centroid.

We note that the caustic crossing effects discussed above differ from the widely known phenomenon of stellar microlensing of lensed quasars. On account of the presence of massive halos, the caustic structure in stellar microlensing is much larger than the BLR, resulting in long-timescale, significant distortions of line profiles \citep{Hutsemekers2023}. By contrast, in the microlensing of IMBH with star clusters, the BLR is larger than the caustics, which leads to statistical fluctuations on short timescales. Differences from microlensing of isolated stellar clusters will be discussed in Section~\ref{sec:contaminant}.

Finally, we use Monte Carlo simulations to predict the probability distribution of the microlensing variability amplitude $\mathcal{A}$ in the presence of stellar clusters. Treating the problem in Einstein radius units renders the results readily applicable to various configurations using the scaling relation in Equation~\ref{sec:magnitude_variation}. We simulate 1000 random spatial configurations of IMBH and star clusters and calculate their magnification maps (Figure~\ref{fig:caustics}b). Finite source effects of the quasar accretion disk are included by convolving the magnification map with the disk surface brightness profile as before. 

We calculate the distribution of $\mathcal{A}$ from the gradient of the magnification map. Denoting the magnification map in $u$-space as ${A}(u)$, the variability amplitude is given by

\begin{equation}
\mathcal{A} = \frac{\nabla {A(u)}}{A} \cdot \frac{\mathrm{d}\vec{u}}{\mathrm{d}t} = \left|\frac{\nabla {A}}{A} \right| \frac{\mathrm{d}{u}}{\mathrm{d}t}\cos\theta ,
\end{equation}

\noindent
where $\nabla A(u)$ is the gradient of the magnification rate, and $\theta$ is the relative angle between the relative velocity vector and the direction of magnification gradient. The factors $\nabla A/A$, $\mathrm{d}{u}/\mathrm{d}t$, and $\cos\theta$ are independent variables and thus sampled separately. The distribution of $\nabla A/A$ is obtained by uniformly sampling locations in 1000 caustic simulations across the two-dimensional $u$-space for $u \leq 1$. The velocity term $\mathrm{d}{u}/\mathrm{d}t$ follows an exponential distribution with a mean of $10^{-3}$\,yr$^{-1}$ (Section~\ref{sec:magnitude_variation}); results for different mean velocities can be derived using the scaling relation in Equation~\ref{eq:scale}. The angle $\theta$ is uniformly distributed. 

The resulting probability distribution of $\mathcal{A}$ shows that, as expected, the microlensing variability is pronounced in the presence of stellar clusters (Figure~\ref{fig:variability_and_magnification}a). High-variability events ($\mathcal{A}>0.01$\,mag\,yr$^{-1}$) become more frequent and are not restricted to regions with extremely small $u$, unlike the single-lens case. We further select subsamples with variability thresholds of $\mathcal{A}\geq0.02\,$mag\,yr$^{-1}$ and 0.01\,mag\,yr$^{-1}$ in Figure~\ref{fig:variability_and_magnification}b. Our results indicate that the average magnification rate of highly variable events is lower than that of single lens systems. Moreover, the average magnification is less sensitive to the threshold of variability (Figure~\ref{fig:mean_magnification}). The relatively low magnification is because the mean magnification rate is dominated by the lensing of the IMBH, while the variability is dominated by the caustic effects of stars. In the presence of caustics, quasars far from the center may still have significant microlensing variability during caustic crossing, in spite of the weak IMBH lensing magnification there.

\section{Distinguishing the Microlensing Signal}
\label{sec:multiband}

This section outlines key observational signatures to differentiate IMBH microlensing from intrinsic quasar variability, including achromaticity, BLR magnification behavior, and deviations from established AGN scaling relations. We focus on long-term quasi-linear microlensing variability, as rare caustic-crossing events are more readily identifiable.

\subsection{Coherent Variability in the Ultraviolet, Optical, and X-rays}
\label{sec: simutaneous}

Distinct from chromatic intrinsic quasar variability \citep[e.g.,][]{Cackett2007MNRAS, Vanden2004ApJ, Kimura_2020_ApJ}, microlensing produces wavelength-independent magnification and variability, synchronized across all bands for point sources. We assess this achromaticity while considering finite source effects.

Microlensing magnification and variability are achromatic when the source size is much smaller than the Einstein radius. Denoting ${\Delta A}$ and ${\Delta \mathcal{A}}$ as the differences in magnification and variability, respectively, between quasar regions separated by $\Delta u$, Equations~\ref{eq:magnification} and \ref{eq:variability} imply ${\Delta A}/{A} \lesssim \Delta u/u$ and ${\Delta \mathcal{A}}/{\mathcal{A}} \lesssim \Delta u/u$. Given typical Einstein radii of $r_{\rm E} \approx 0.4$\,pc (Figure~\ref{fig:thetaE}b) and accretion disk sizes of $r_s \approx 10^{-2}$\,pc \citep{du_2016, li_2021}, we have $\Delta u = r_s/r_{\rm E} \approx 0.02$. For a typical $u \approx 0.2$ (Figure~\ref{fig:variability_and_magnification}b), the magnification and variability across different parts of the quasar accretion disk can deviate by only $\sim$10\%. The differences become significant in rare cases where $u < 0.02$, but such events are identifiable by their extremely high magnification. Hence, variations in magnification and variability across the accretion disk are generally negligible. Since the ultraviolet and optical continuum emission of quasars is dominated by the accretion disk, and X-ray emission originates from the hot corona near the inner disk, we expect to observe almost identical lensing magnification and variability across the X-ray, ultraviolet, and optical bands. By comparison, microlensing by stars has much smaller Einstein radii and is chromatic due to temperature gradients in the accretion disk \citep{Jimenez2014ApJ}.

Microlensing-induced variability is synchronous across all wavelengths. The arrival time difference between microlensing images follows \citep{mao_gravitational_1992}

\begin{equation}
\Delta t \approx 2 \times 10^{-5} (1+z_l) \left({M}/{M_\odot}\right)\, \mathrm{s},
\end{equation}

\noindent
For IMBHs of $M = 10^3\,M_\odot$, $\Delta t\lesssim 0.02$\,s, which is observationally negligible. Similarly, lensing-induced time delays between different regions of the quasar are also of order $\Delta t$. Therefore, microlensing can be considered effectively simultaneous between different lensing images and quasar regions, and the microlensing light curves are synchronized across different bands. In contrast, intrinsic quasar variability exhibits time lags within the accretion disk due to outward-propagating thermal fluctuations \citep[e.g.,][]{Jiang2017ApJ, son_2025}, and delays of days to months between the accretion disk and the BLR arise from their spatial separation \citep{kaspi_relationship_2005, du_2016}.
}

\subsection{Magnification of Emission-line Signal}
\label{sec:blr}

Quasar emission lines in the optical and ultraviolet band are mostly produced by the BLR and the narrow-line region. The two regions differ not only in their kinematics but also structure and location in the galaxy. The BLR is located close to the accretion disk of the AGN, with an outer edge delineated by the dusty and molecular torus (e.g., \citealt{czerny_origin_2011}). The strong emission lines, broadened largely by gravity and powered by photoionization by the continuum from the accretion disk.

The exact value of the magnification rate of the BLR may differ slightly from that of the accretion disk. We have demonstrated that when the IMBH is much more distant than the substructures of the quasar, the different parts of the quasar can be viewed as a single point source. Meanwhile, as discussed in Section~\ref{sec:magnitude_variation}, microlensing events with large variation rates usually correspond to a small $u$ (i.e., a small distance between the IMBH and the centroid of the quasar). As a result, the magnification rate of the BLR may be slightly different from that of the accretion disk if the IMBH is too close. 

To quantify the difference between the BLR and the accretion disk, we adopt the model of \cite{li_bayesian_2013}, which parameterizes the radial distribution of the line-emitting clouds as

\begin{equation}
R = fr +(1-f) \mathcal{R},
\end{equation}

\noindent
where $r$ is the mean radius of the BLR, $f$ is a model parameter, and $\mathcal{R}$ is a random variable following the $\Gamma$ distribution with a mean of $r$ and a standard deviation of $\beta r$. We consider two scenarios for $r = 0.02\,$pc and $ 0.05\,$pc, fixing the other parameters to the typical values in \cite{li_bayesian_2013}\footnote{Parameters include $\beta = 0.5$, $f=0.5$, an opening angle of the BLR clouds of $50^\circ$, a viewing angle of $45^\circ$, and the non-linearity factor of the BLR response rate as $-0.2$.}. With this setup, we simulate $10^4$ clouds and derive their luminosity and spatial distribution projected along the line-of-sight. Placing the IMBH randomly within the Einstein radius, for which we assume a typical value of $r_\mathrm{E} = 0.4$\,pc, we calculate the total magnification rate of the BLR as a function of the dimensionless angular separation $u$ between the IMBH and the quasar.

\begin{figure}
\centering
\includegraphics[width=0.98\linewidth]{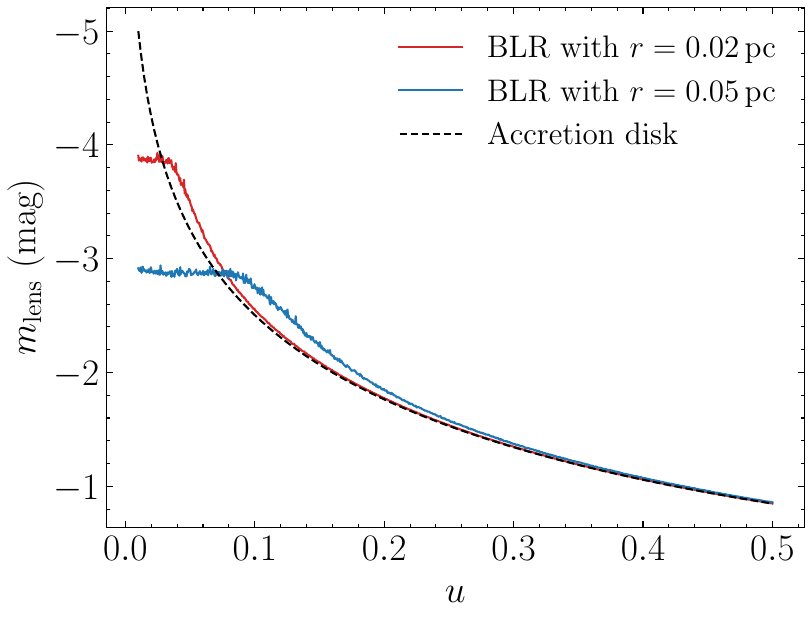}
\caption{Microlensing magnified magnitude $m_\mathrm{lens}$ of the BLR with mean radius $r=0.02$\,pc and $r=0.05$\,pc for different dimensionless angular separation $u$ between the IMBH and the quasar. For comparison, the dashed line shows the microlensing magnified magnitude of the quasar accretion disk.}
\label{fig:BLR}
\end{figure}

Figure~\ref{fig:BLR} shows that as $u$ gets smaller the magnification rate of the BLR initially increases and then saturates upon reaching a turning point $u\approx r/r_\mathrm{E}$. When $u\gtrsim r/r_\mathrm{E}$, the magnification rate of the BLR closely follows that of the accretion disk to within 0.2\,mag. In this situation, the continuum and broad-line emission both appear extraordinarily luminous, but their flux ratio remains constant. When $u\lesssim r/r_\mathrm{E}$, although the BLR is still highly magnified, its magnification rate is much smaller than that of the accretion disk, and we should observe an abnormally low flux ratio of the broad-line emission to the continuum. Moreover, since the magnification rate of the BLR is insensitive to the position of the IMBH, the relative motion between the quasar and the IMBH will not cause variation of the magnification rate of the BLR. Thus, we will only observe variability in the continuum flux, not in the broad emission lines.

In contrast to the BLR, the narrow-line region is located far from the accretion disk and outside the dusty torus, with distances $\gtrsim 10\,$pc \citep{hickox_obscured_2018}, much larger than the Einstein radius. Under these circumstances, the microlensing of the IMBH induces almost no magnification on the narrow emission lines. The highly magnified accretion disk and BLR would produce a quasar spectrum with a bright nonstellar continuum superposed with prominent broad emission features but weak narrow lines.

\subsection{Extraordinarily Luminous Outliers in Scaling Relations}
\label{sec:outliers}

It has been well-established that the size of the BLR is tightly correlated with the luminosity of the quasar with a power-law relation $R_\mathrm{BLR} \propto L^\alpha$. The power index $\alpha$ is 0.53 for the optical continuum luminosity, with an intrinsic scatter of only $13\%$ \citep{bentz_2013}, despite mild changes in the slope and increased scatter for AGNs of high Eddington ratios \citep{du_2016}. The size-luminosity relation also extends to other wavelengths, including the ultraviolet and X-ray continuum luminosities and the H$\beta$ line luminosity, although with different power indices and slightly larger scatter \citep{kaspi_relationship_2005}. The size of the BLR is usually measured with the reverberation mapping method \citep{blandford_1982} using the time lag between the continuum and the broad-line emission. Because microlensing caused by an IMBH will not introduce any noticeable time lag between the different parts and different images of the AGN (Section~\ref{sec: simutaneous}), the BLR size measured using reverberation mapping is not affected. In contrast, the optical, ultraviolet, X-ray, and broad H$\beta$ luminosity are highly magnified. As discussed in Section~\ref{sec:magnitude_variation}, high yearly magnification rate usually corresponds to high total magnification rate. For instance, a quasar with a yearly magnification rate higher than 0.02$\,\mathrm{mag\,yr^{-1}}$ typically has a magnification rate on the optical continuum $\gtrsim 3\,$mag. These AGNs will appear to have an exceptionally compact BLR for their luminosity, with $R_\mathrm{BLR}$ smaller by a factor of $\sim 4$ relative to the scaling relation, far beyond the intrinsic scatter, mimicking the shortened lags characteristic of super-Eddington sources \citep{du_2016} but unlike them in their spectral properties (e.g., lacking strong Fe~II emission; \citealt{dong_2011}). 

In cases of extremely high magnification, with rates $\gtrsim 4\,$mag, the BLR magnification will differ from that of the accretion disk. This difference may be sufficient to make a lensed quasar deviate from the correlation between the 5100$\,\text{\AA}$ continuum luminosity and the luminosity of the broad H$\alpha$ or H$\beta$ line, which has an intrinsic scatter of only 0.2 dex \citep{greene_estimating_2005}. AGNs with high magnification are inherently more luminous and exhibit greater microlensing variability, making them more easily detectable; therefore, the fraction of high-magnification cases may be considerable. As shown in Figure~\ref{fig:BLR}, the magnification rate of the BLR is $\sim 1\,$mag lower than that of the accretion disk, with the exact difference depending on the BLR size and the separation between the IMBH and the quasar, or, equivalently, the total magnification rate. We expect lensed quasars to deviate from the optical continuum versus Balmer line correlations by $\sim 1.5\,\sigma$, another feature that can be exploited to identify microlensing candidates.

Finally, microlensing by IMBHs also causes quasars to deviate from the empirical relation between mid-infrared and X-ray luminosity, in which the 12\,$\mu$m luminosity is almost linearly proportional to the 2--10\,keV luminosity with an intrinsic scatter of $\sim0.3$ dex \citep{asmus_subarcsecond_2015}. While the X-ray emission of a quasar lensed by an IMBH will be significantly amplified because it comes from the corona around the accretion disk, the mid-infrared emission from the dusty torus will not be magnified considering that its size is much larger than the Einstein radius of the IMBH. For a magnification rate of 3\,mag, the lensed quasar will lie $4\,\sigma$ above the intrinsic scatter of the mid-infrared versus X-ray correlation.

{
\subsection{Comparison with Contaminant Sources}
\label{sec:contaminant}

We presented critical properties of IMBH microlensing. In this section, we discuss the characteristics of possible contaminant sources for comparison. These contaminants may include intrinsic quasar variability, stellar microlensing, microlensing by dark matter subhalos, and variability caused by clouds moving across the line-of-sight.

Quasars are known to show stochastic variability on time scales of several months \citep[e.g.,][]{sanchez_2019,suberlak_improving_2021}. This significantly shorter time scale makes it less likely to be confused with the long-term microlensing signals, provided a sufficient observation time span. However, stochastic variability can still hamper the detection of the long-term microlensing signals by introducing noise that lowers the detection limit. We analyze its impact on microlensing detection in Section~\ref{sec:detectability}.

Quasars may also have long-term variability as they ignite and fade over their lifetime. This process can be traced by extended emission-line regions, which encode the quasar emission from thousands of years earlier \citep{Keel2017ApJ}. This long-term variability owing to, for instance, state transitions in AGN accretion disks, would be wavelength-dependent because it is accompanied by temperature changes \citep{Sartori_2018_MNRAS, Li2024}. It is well-documented that quasars become bluer as they brighten \citep{Schmidt2012ApJ, Ruan2014ApJ}. Moreover, this variability is not strictly synchronized across different regions of the quasar, resulting in variations across different wavelengths \citep{Sun2014ApJ}. These properties contrast with the achromatic and synchronized nature of microlensing-induced long-term variability.

Similarly, absorption from line-of-sight clouds may imprint detectable reddening in the optical/ultraviolet, as well as photoelectric absorption at X-ray energies. The variability resulting from cloud movement and changes in obscuration is more pronounced at bluer wavelengths. This type of variability can be easily distinguished from microlensing effects through multiband observations.

Stars in the intervening space can occasionally cause microlensing of quasars. Dense star regions may form caustics that produce brightness fluctuations and even flares in quasars. Such events are more likely to occur when quasars are positioned behind foreground galaxies, which are usually associated with strong lensing. Since quasars are typically observed at high Galactic latitudes and in sparser extragalactic fields, stellar microlensing should be relatively rare. In any case, several strategies can be employed to distinguish stellar microlensing from IMBH microlensing. 

First, stars are much less massive than IMBHs, resulting in a smaller Einstein radius. Although young stars can be massive, stars in galaxy halos or away from galaxies are typically part of an old stellar population, consisting mainly of low-mass stars and stellar remnants \citep{Gallart2019NatAs}. This smaller Einstein radius means that stellar microlensing cannot magnify both the accretion disk and the BLR simultaneously, nor will it yield a similar magnification rate. Second, even if multiple stars were to magnify the accretion disk and the BLR separately, this scenario is still distinguishable from IMBH microlensing. The BLR line width has a significant contribution from the Keplerian motion of the clouds. Since the small Einstein radius of stars can only magnify parts of the BLR, the resulting line flux would be dominated by magnified emission from these localized regions. Consequently, the width of the emission line would appear unusually narrow, as it loses the broader contribution from the Keplerian motions of the entire cloud ensemble. Lastly, the caustics from stellar microlensing also differ from those in IMBH microlensing. The caustics from an IMBH with a star cluster show a centrally symmetric, elongated, diamond-shaped structure, with magnification dominated by the contribution from the central IMBH. By contrast, the caustic network of an isolated star cluster is diffuse, web-like, and irregular, lacking a dominant central feature. We can infer the local caustic patterns from the multi-wavelength properties of the accretion disk and BLR magnification \citep[e.g.,][]{Anguita2008}. Furthermore, stellar microlensing lacks the long-term variability that is characteristic of IMBH microlensing. Between caustic crossing events in stellar microlensing, the quasar typically experiences little to no magnification and thus does not behave as a luminous outlier, as discussed in Section~\ref{sec:outliers}.

Stars in the vicinity of the Milky Way pose a different challenge. At such short distances, the Einstein radius of stars and even less massive objects can exceed that of IMBHs. However, the proximity of Galactic stars introduces significant parallax effects, which modulate the light curves. Additionally, their short distance results in large angular velocities, leading to significantly shorter event time scales and noticeable astrometric microlensing effects.

Stars within host galaxy of the quasar present little confusion. As the Einstein radius decreases with lens distance, the Einstein radius for these stars becomes extremely small---smaller than both the BLR and the accretion disk. These stars can be distinguished using the finite source effects of the quasar accretion disk and the diagnostics discussed in Section \ref{sec:blr}. The small Einstein radius also significantly reduces the event rate, making such occurrences rare.

Dark matter subhalos are much more massive and diffuse than IMBHs, so that they induce smooth, extended lensing effects instead of microlensing. This occurs because their large physical size means that the magnification remains nearly constant as the quasar moves within the gravitational field of the subhalo. Subhalos with masses $\gtrsim10^6\,M_\odot$ can imprint gravitational lensing time delays on timescales of minutes.
}

{

\begin{figure}
\centering
\includegraphics[width=0.98\linewidth]{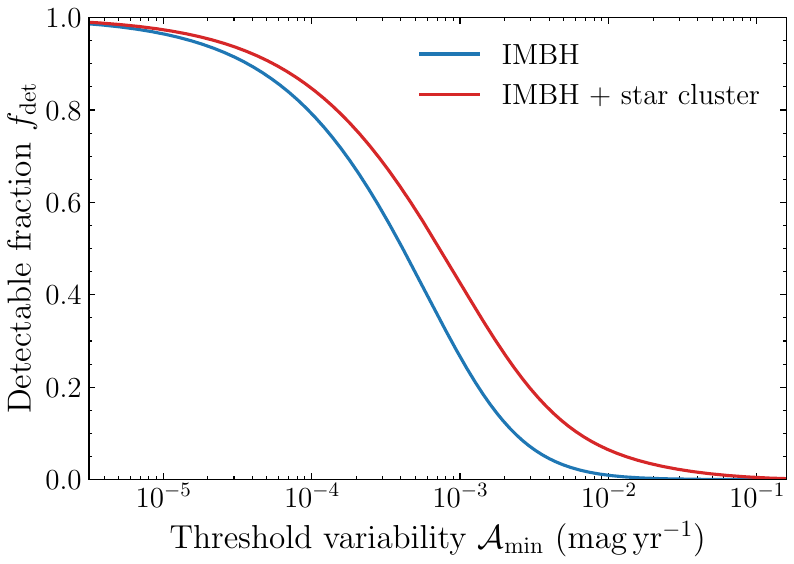}
\caption{The detectable fraction $f_\mathrm{det}$ of microlensing events, defined as the proportion of events with microlensing variability exceeding the threshold $\mathcal{A}_\mathrm{min}$ relative to all events within the Einstein radius. The blue and red curves represent microlensing by an isolated IMBH and an IMBH with a star cluster, respectively.}
\label{fig:cdf_of_variability}
\end{figure}

\section{Event Number}
\label{sec:event_number}

We estimate the number of IMBH microlensing events that will be observable in future surveys. This number primarily depends on three factors: the total number of monitored quasars, the optical depth of IMBH microlensing, and the detectable fraction of such events. 

The detectable fraction $f_\mathrm{det}$ is the proportion of microlensing events with microlensing variability $\mathcal{A}$ above the detection threshold $\mathcal{A}_\mathrm{min}$. Mathematically, this fraction is equivalent to the complement of the cumulative distribution function (CDF) of the variability amplitude $\mathcal{A}$, considering random distribution of relative velocities and locations for $u\leq 1$, such that

\begin{equation}
 f_\mathrm{det}=1-\mathrm{CDF}(\mathcal{A}_\mathrm{min}) .
\end{equation}

\noindent
We calculate CDF of $\mathcal{A}$ by integrating the PDF in Figure~\ref{fig:variability_and_magnification}a and thereby obtain $f_\mathrm{det}$ at various thresholds $\mathcal{A}_\mathrm{min}$ (Figure~\ref{fig:cdf_of_variability}) for isolated IMBHs (Section~\ref{sec:magnitude_variation}) and IMBHs embedded in star clusters (Section~\ref{sec:caustics}). For $\mathcal{A}_\mathrm{min}=0.01$\,mag\,yr$^{-1}$, $f_\mathrm{det}=0.01$ for isolated IMBHs and $f_\mathrm{det}=0.07$ for IMBHs with star clusters. IMBHs with star clusters have a higher detectable fraction because the concentration of stars produce caustic structures that amplify microlensing variability by creating sharp magnification gradients (Section~\ref{sec:caustics}). As mentioned in Section~\ref{sec:magnitude_variation}, the PDF of $\mathcal{A}$ is for quasars at a typical redshift $z_s=2$. However, our results are readily generalizable to other redshifts using the scaling relation $\mathcal{A}\propto 1/(1+z_s)r_\mathrm{E}$ (Equation~\ref{eq:scale}), where $r_\mathrm{E}(z_s)$ is the Einstein radius in the quasar plane. Specifically, results of $f_\mathrm{det}$ at redshift $z_s$ can be obtained by rescaling $\mathcal{A}_\mathrm{min}$ in Figure~\ref{fig:cdf_of_variability} by the ratio of $(1+z_s)r_\mathrm{E}$ to that at $z_s=1$.

The optical depth refers to the probability for a source being within the Einstein radius of any lensing object along the line-of-sight \citep{schneider_gravitational_1992}. For a population of lensing IMBHs with a total comoving mass density $\rho_\mathrm{IMBH}$, the optical depth for a quasar at redshift $z_s$ is given by \citep{munoz_lensing_2016} 

\begin{equation}
\begin{aligned}
\tau(z_s) &= \int \mathrm{d} M\int_0^{D_s} \mathrm{d}D_l \, \pi \theta_\mathrm{E}^2 D_l^2 n_{\mathrm{IMBH}} \\
&= \frac{4\pi G}{c^2} \rho_{\mathrm{IMBH}} \int_0^{D_s} \mathrm{d}D_l \, (1+z_l) \frac{D_l (D_s - D_l)}{D_s},
\end{aligned}
\label{eq:optical_depth }
\end{equation}

\noindent
where $D_l$ and $D_s$ are the comoving distances to the IMBH and quasar, respectively, $n_\mathrm{IMBH}$ is the comoving IMBH number density, and $\theta_\mathrm{E}$ is Einstein radius in Equation~\ref{eq:einstein_radii}. The second equality uses $\rho_\mathrm{IMBH}=\int M n_\mathrm{IMBH}(M)dM$. We adopt IMBH mass density $\rho_\mathrm{IMBH}=5\times 10^7\,M_\odot\,$Mpc$^{-3}$, which \cite{paynter_evidence_2021} derive from a gravitationally lensed gamma-ray burst, to calculate the optical depth as a function of quasar redshift $z_s$ (Figure~\ref{fig:optical_depth}a). As $\tau\propto\rho_\mathrm{IMBH}$, results for different $\rho_\mathrm{IMBH}$ scale accordingly. We further compute the ``detectable'' optical depth $\tau f_\mathrm{det}$, which represents the probability that a quasar is lensed and exhibits detectable microlensing variability, for a range of redshift $z_s$ and threshold $\mathcal{A}_\mathrm{min}$. Figure~\ref{fig:optical_depth}b presnets the results for $\mathcal{A}_\mathrm{min}=0.01$\,mag\,yr$^{-1}$. Our results indicate that one in a million quasars at $z_s=1$ can produce significant lensing variability signals with $\mathcal{A}\geq0.01$\,mag\,yr$^{-1}$; this fraction is $\sim 10$ times higher if the IMBH resides in a star cluster.

\begin{figure}
\centering
\includegraphics[width=0.98\linewidth]{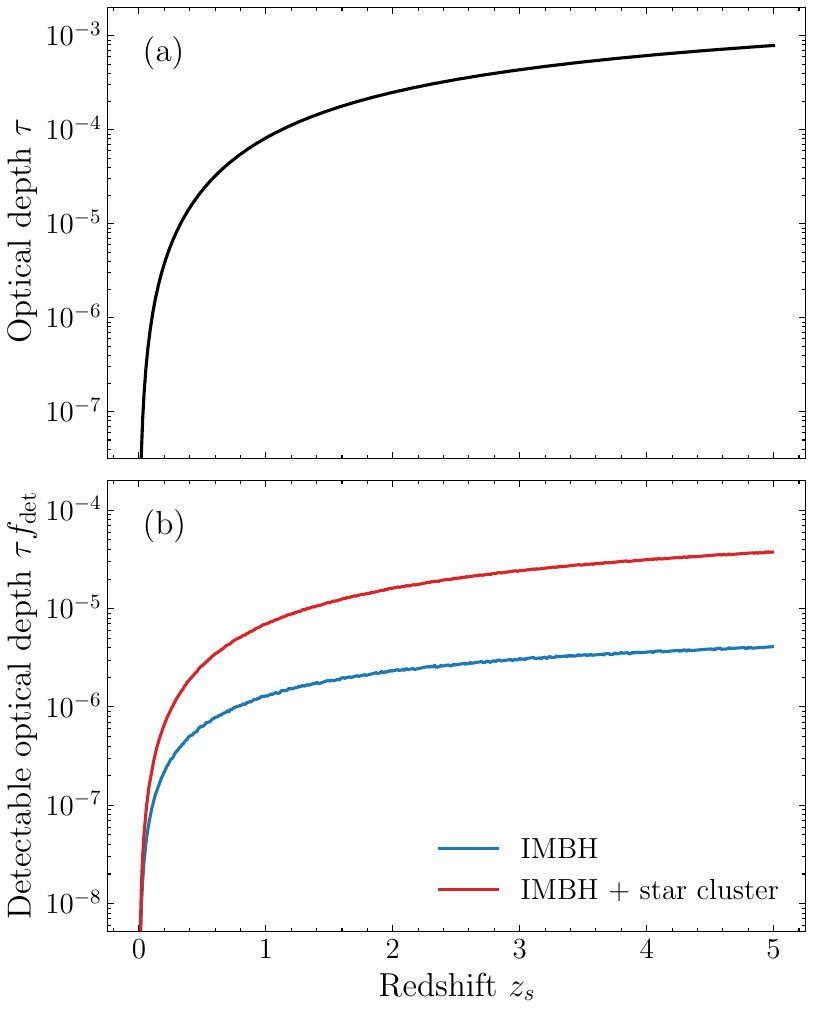}
\caption{(a) The optical depth $\tau$ for IMBH microlensing, indicating the probability that a quasar at redshift $z_s$ lies within the Einstein radius of any intervening IMBHs. (b) The detectable microlensing optical depth $\tau f_\mathrm{cal}$, which is the product of the optical depth $\tau$ and the detectable fraction $f_\mathrm{cal}$, represents the probability of a quasar undergoing lensing and showing detectable variability. Here the detection limit for variability is $0.01\,\mathrm{mag\,yr}^{-1}$. The blue and red curves represent microlensing by an isolated IMBH and an IMBH surrounded by a star cluster, respectively.}
\label{fig:optical_depth}
\end{figure}

Finally, we calculate the quasar number using the evolving luminosity function of quasars $\phi_{\mathrm{bol}}(L,z_s)$ from \cite{shen_bolometric_2020}, for quasars with observed $g$-band magnitude brighter than 25\,mag, the detection limit of LSST \citep{ivezic_lsst_2019}. Integrating $\phi_{\mathrm{bol}}$ over bolometric luminosity $L$ above threshold $L_\mathrm{min}(z_s)$, the observable number of quasars at redshift $z_s$

\begin{equation}
N_\mathrm{quasar}(z_s) \, \mathrm{d}z_s = \mathrm{d}z_s \, \frac{\mathrm{d}V_c}{\mathrm{d}z_s} \int_{\log L_\mathrm{min}}^{\infty} \phi_\mathrm{bol} \, \mathrm{d}\log L,
\end{equation}

\noindent
where ${\mathrm{d}V_c}/{\mathrm{d}z_s}$ is the differential comoving volume. The threshold $L_\mathrm{min}$ is determined by the observed magnitude limit and is derived considering the luminosity distance, bolometric correction $\kappa_g=-3$ \citep{shen_bolometric_2020}, and a $K$-correction derived using the mean quasar spectral energy distribution template from \cite{shen_bolometric_2020} and the $g$-band transmission profile from the SVO Filter Profile Service\footnote{\tt{http://svo2.cab.inta-csic.es/theory/fps}.}. We also consider lensing magnification effects that lowers the required $L_\mathrm{min}$. We adopt the average lensing magnification $m_\mathrm{lens}$ from Figure~\ref{fig:mean_magnification} in calculating $L_\mathrm{min}$ under variability detection thresholds $\mathcal{A}_\mathrm{min}$ ranging from $10^{-3}$ to $10^{-1}$\,mag\,yr$^{-1}$, for IMBHs with and without star clusters, respectively.

The event number
\begin{equation}
N = \int \mathrm{d}z_s \, N_\mathrm{quasar} \tau f_\mathrm{det}
\end{equation}

\noindent
for different detection thresholds of the variability amplitude $\mathcal{A}_\mathrm{min}$ (Figure~\ref{fig:detectable_events}) indicates that LSST may detect substantial IMBH microlensing events given the IMBH mass density estimated in \cite{paynter_evidence_2021}. The detectable event number is sensitive to the variability detection threshold, especially for isolated IMBHs. Most events come from redshift 1 to 3 (Figure~\ref{fig:detectable_events_z}) because the cosmological volume reaches the largest for both quasars and IMBHs, while at the same time not placing the quasars out of detectable reach.

\begin{figure}
\centering
\includegraphics[width=0.98\linewidth]{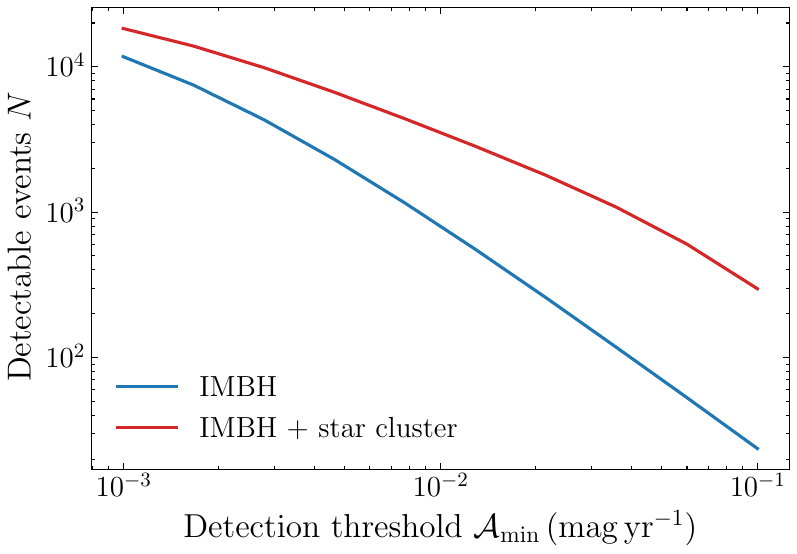}
\caption{The estimated number of IMBH microlensing events $N$ detectable in the LSST survey as a function of the variability detection threshold $\mathcal{A}_\mathrm{min}$. The blue and red curves represent microlensing by an isolated IMBH and an IMBH surrounded by a star cluster, respectively. A $3\,\sigma$ detection threshold for typical quasar variability over the 10-year LSST survey is 0.07\,mag\,yr$^{-1}$ (Section~\ref{sec:detectability}).}
\label{fig:detectable_events}
\end{figure}

\begin{figure}
\centering
\includegraphics[width=0.98\linewidth]{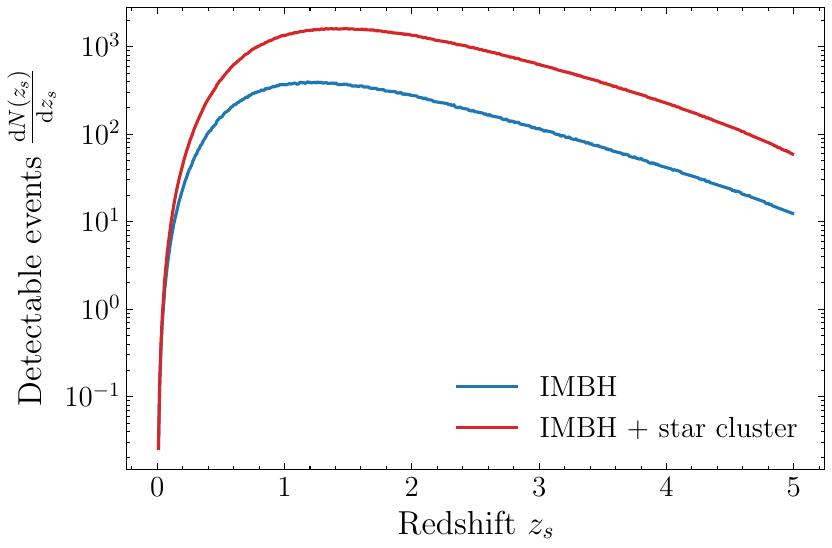}
\caption{The redshift distribution of quasars with detectable microlensing variability, assuming a variability detection threshold of $0.01\,\mathrm{mag\,yr}^{-1}$. The blue and red curves represent microlensing by an isolated IMBH and an IMBH surrounded by a star cluster, respectively.}
\label{fig:detectable_events_z}
\end{figure}

The detectable event number is influenced by several other factors:

\begin{enumerate}
\item We assume the same detection threshold for all quasars, while in practice less luminous quasars may suffer contamination from their host galaxies, which dilute the variability by a factor $L_\mathrm{AGN}/(L_\mathrm{AGN}+L_\mathrm{galaxy})$. Moreover, as their magnitude approaches the photometric limit, microlensing variability is more difficult to detect because of photometric noise. Considering that less luminous quasars are more abundant, the detectable events would largely depend on the detection threshold for faint quasars.

\item Our calculations adopt an IMBH mass of $10^3\,M_\odot$. While the IMBH optical depth remains unaffected by the choice of mass, as it depends only on the total IMBH mass density, the variability amplitude $\mathcal{A}$ scales with mass (Equation~\ref{eq:scale}). However, the dependence is weak, for $\mathcal{A}\propto1/\sqrt{M}$. The event number resulting from choosing a different mass would agree within an order of magnitude for a fixed detection threshold.

\item The event number scales with the assumed IMBH mass density, which, unfortunately, varies significantly from study to study. We adopt $\rho_\mathrm{IMBH}=5\times 10^7\,M_\odot\,$Mpc$^{-3}$ from \cite{paynter_evidence_2021}. Observations of hyperluminous X-ray sources instead suggest $\rho_\mathrm{IMBH} \approx 10^6\,M_\odot\,$Mpc$^{-3}$ for IMBHs akin to those purported to exist in HLX-1 and M82 X-1 \citep{caputo_estimating_2017}. Illustris TNG50 simulations suggest that Milky Way-like galaxies host thousands of wandering IMBHs with masses of $10^{4}-10^{5}\,M_\odot$ \citep{weller_dynamics_2022}, corresponding to $\rho_\mathrm{IMBH} = 10^6-10^7\,M_\odot$\,Mpc$^{-3}$ when scaled to the cosmological abundance of galaxies with similar masses \citep{Kelvin2014MNRAS}. Theories of primordial black holes predict $\rho_\mathrm{IMBH} \approx 10^7M_\odot$\,Mpc$^{-3}$ in the IMBH mass range \citep{Garca_Bellido_2017}. Estimates based on ultracompact dwarf galaxy counts yield a significantly lower IMBH density of $\sim10^{3}\,M_\odot$\,Mpc$^{-3}$ \citep{greene_intermediate-mass_2020}. Although these estimates span several orders of magnitude, many suggest that a promising number of IMBH microlensing events could be detected. Importantly, even a null detection would place meaningful constraints on the cosmological mass density of IMBHs, thereby testing different formation scenarios.
\end{enumerate}

\section{Detection Methodology}
\label{sec:detectability}

As the detectable number of events increases significantly with the detection limit, it is critical to enhance our toolkit to detect long-term lensing signals with advanced statistical methods. This section analyzes noise sources influencing signal detection, develops a framework for detecting lensing signals, predicts detection limits under various observational conditions, and validates our results using simulated data.

\subsection{Noise Budget}

The detection of microlensing variability is affected by two types of noise: time-independent white noise from photometry and time-correlated red noise from quasar intrinsic variability. Modern image differencing techniques (e.g., \citealt{bramich_new_2008, wozniak_crowded_2008}) suppress photometric noise effectively. For example, \cite{Aihara_2022_PASJ} reported photometric errors of $\sim0.06\,\mu$Jy (corresponding to 0.01--0.17\,mag for $i=22$--25\,mag AGNs) in Subaru observations resembling LSST conditions \citep{Kimura_2020_ApJ}

Intrinsic quasar variability dominates the noise budget. This variability is thought to arise from accretion disk temperature fluctuations \citep{kelly_are_2009}, and is modeled as a damped random walk (DRW) with typical correlation time $\tau\sim200$\,d and amplitude $\sigma\approx0.2$\,mag \citep{suberlak_improving_2021}. Studies of OGLE quasar light curves confirm that the DRW model accounts for $\gtrsim90\%$ of variability at $\sim$10\% photometric precision \citep{zu_is_2013, udalski_ogle-iv_2015}. While deviations may occur at $\sim2$\,hr timescales \citep{Kasliwal2015MNRAS}, these are irrelevant for LSST’s daily cadence and IMBH microlensing extensive timescales.

\begin{figure}
\includegraphics[width=0.98\linewidth]{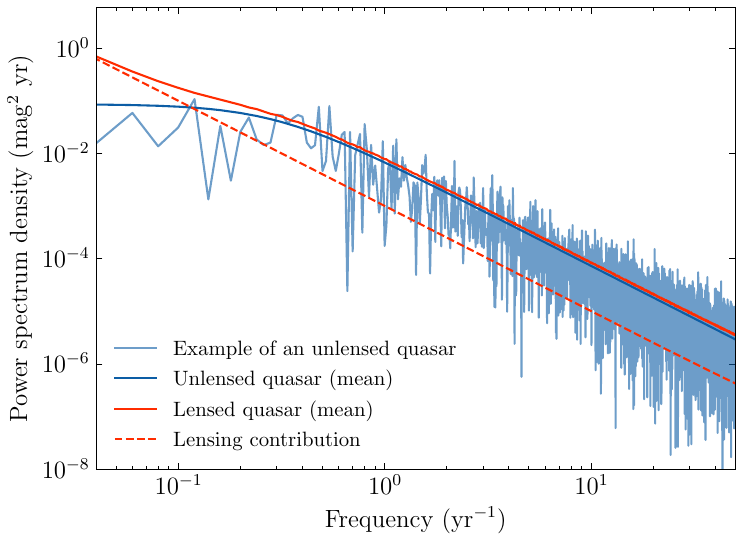}
\caption{Power spectral density (PSD) of simulated quasars. The shallow blue curve plots the PSD of a simulated unlensed quasar with $\tau=200$\,d, $\sigma=0.2$\,mag, while the dark blue curve shows the average PSD of 100 realizations. The red solid curve gives the PSD of lensed quasars with lensing amplitude of $0.02$\,mag\,yr$^{-1}$, and the red dashed line shows the PSD of the lensing signal alone. The lensing signal dominates at low frequencies, while quasar variability dominates at high frequencies.}
\label{fig:psd}
\end{figure}

\begin{figure*}
\centering
\includegraphics[width=0.98\textwidth]{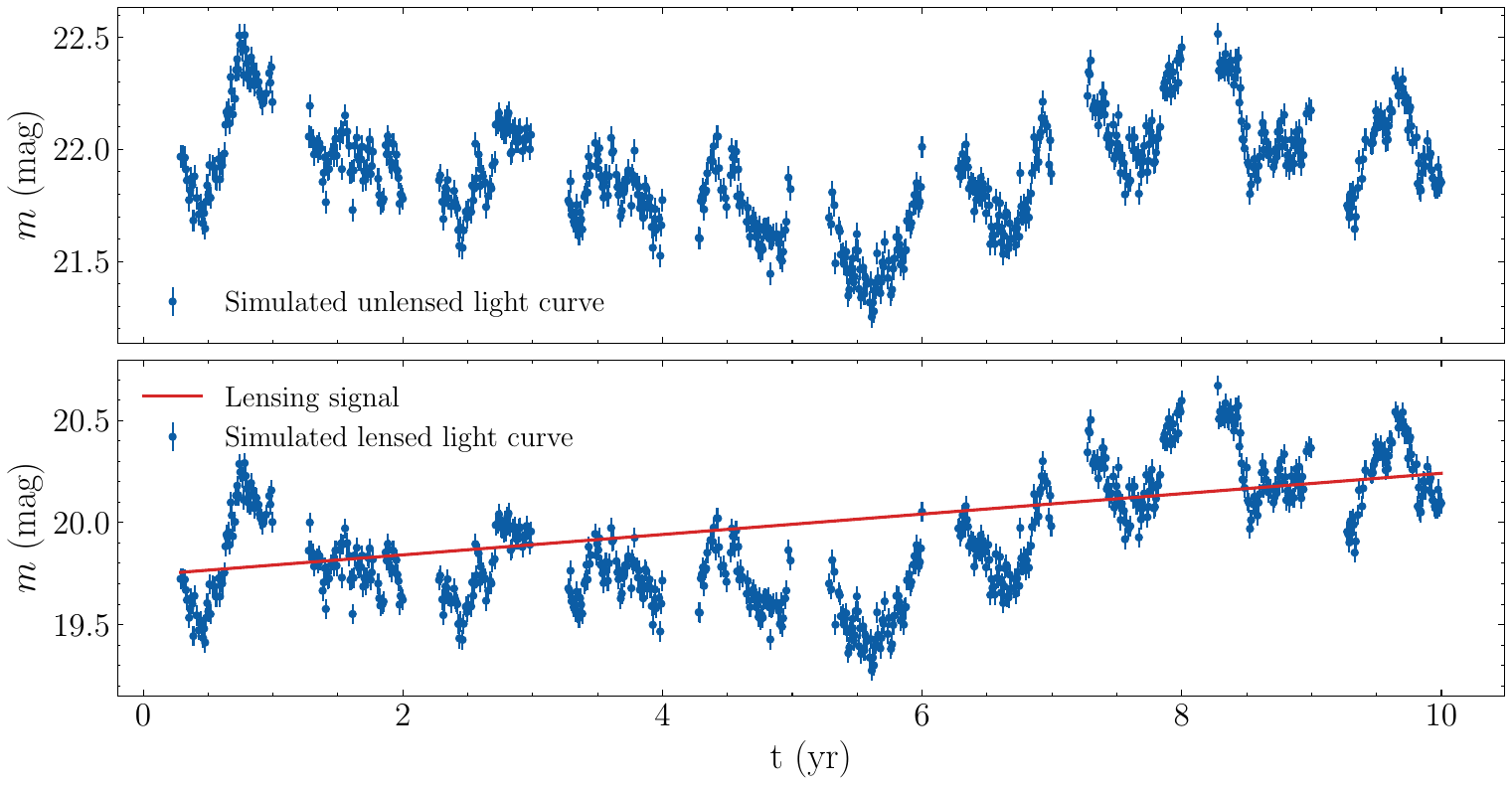}
\caption{Simulated light curve of a quasar without (top) and with (bottom) a microlensing signal (red line) of amplitude $\mathcal{A}=0.05\,$mag\,yr$^{-1}$. The light curves are sampled every 3 days over a 10-year period, incorporating seasonal gaps and occasional missing data. We add photometric noise with $\sigma_0=0.05\,$mag. The light curve is simulated with DRW parameters $\tau=200\,$d and $\sigma=0.2\,$mag. The unlensed quasar has a mean magnitude of 22, while the lensed quasar has a brighter mean magnitude of 20 due to the magnification effect of gravitational lensing.}
\label{fig:simulated_light_curve}
\end{figure*}

\subsection{Analytical Prediction of the Detection Limit}

To detect microlensing signals amidst red noise, we employ the matched-filter technique \citep{creighton_2011_book}, optimal for extracting known signals in noise with characterized correlations. This method, used successfully in gravitational wave detection and exoplanet transit studies \citep{ruffio_improving_2017, robnik_matched_2021}, maximizes the signal-to-noise ratio (SNR) by weighting frequencies inversely by their noise power.

We model the magnitude variability as

\begin{equation}
m(t) = m_0 + n(t) + \mathcal{A}\,t,
\end{equation}

\noindent
where $m_0$ is the quasar magnitude at $t=0$, $n(t)$ is DRW noise with time scale $\tau$ and amplitude $\sigma$, and $\mathcal{A}\,t$ is the microlensing linear variability with linear slope $\mathcal{A}$. The definition of $\mathcal{A}$ here is consistent with that in Equation~\ref{eq:variability}, considering the linear approximation and $-2.5\log A\approx \ln A$. 

The detection problem is thus to measure $\mathcal{A}$ optimally amidst the DRW noise $n(t)$. The DRW power spectral density \citep{kelly_are_2009} is

\begin{equation}
S_{n}(f) = \frac{4\sigma^2\tau}{1+(2\pi \tau f)^2}.
\label{eq:power_spectrum}
\end{equation}

\noindent
Under the null hypothesis $\mathcal{H}_0$ of no lensing signals, the likelihood of observing $s(t)$ is

\begin{equation}
P(s|\mathcal{H}_0) \propto e^{-\frac{1}{2}\langle s,s \rangle},
\label{eq:log_like_H0}
\end{equation}

\noindent
which is similar to the formalism in time-independent Gaussian noise, despite that the inner product here is defined in Fourier space, weighted by the noise power spectrum. Following the formalism in \cite{creighton_2011_book},

\begin{equation}
\langle s,s \rangle \equiv \int \frac{|\tilde{s}(f)|^2}{S_{n}(f)/2} \mathrm{d}f,
\label{eq:inner_product}
\end{equation}

\noindent
with $\tilde{s}(f)$ the Fourier transform of ${s}(t)$. For the alternative hypothesis $\mathcal{H}_1$ that lensing is present with amplitude $\mathcal{A}$, the likelihood of observing $s(t)$ becomes

\begin{equation}
P(s|\mathcal{H}_1) \propto e^{-\frac{1}{2}\langle s-\mathcal{A}t, s-\mathcal{A}t\rangle}.
\label{eq:log_like_H1}
\end{equation}

The best-estimate of $\mathcal{A}$ corresponds to that which maximizes the ratio $P(s|\mathcal{H}_1)/P(s|\mathcal{H}_0)$. According to \cite{creighton_2011_book}, the solution is $\hat{\mathcal{A}} = \langle s(t),t\rangle/\langle t,t\rangle $ and the uncertainty of the estimation is $\sigma_{\hat{\mathcal{A}}} = 1/\sqrt{\langle t,t\rangle}$. We calculate the inner product using Equations~\ref{eq:power_spectrum} and \ref{eq:inner_product}. In the limit that the observation time span $T\gg \tau$, we obtain, after some calculations, 

\begin{equation}
\sigma_{\hat{\mathcal{A}}} \approx 2\sqrt{6}\,\sigma\,\tau^{1/2}\,T^{-3/2}.
\label{eq:analytical}
\end{equation}

\noindent
Details of all the derivations are in Appendix~\ref{appendix:detection}. This expression indicates that the sensitivity to the lensing variability dramatically increases with observation time span, and decreases with the timescale and amplitude of quasar intrinsic variability. Moreover, as we discuss in Appendix~\ref{appendix:detection}, the detection is not sensitive to the cadence of the observation, as long as it is significantly shorter than $\tau$.

For a 10-year LSST-like survey monitoring quasars with $\sigma=0.2$\,mag and $\tau=200$\,d, Equation~\ref{eq:analytical} gives variability amplitude uncertainty of 0.023\,mag\,yr$^{-1}$. Extending to 20 years improves this to $0.008$\,mag\,yr$^{-1}$.

}

\subsection{Detection Limit on the Simulated Data}

\begin{figure*}
\centering
\includegraphics[width=0.9\linewidth]{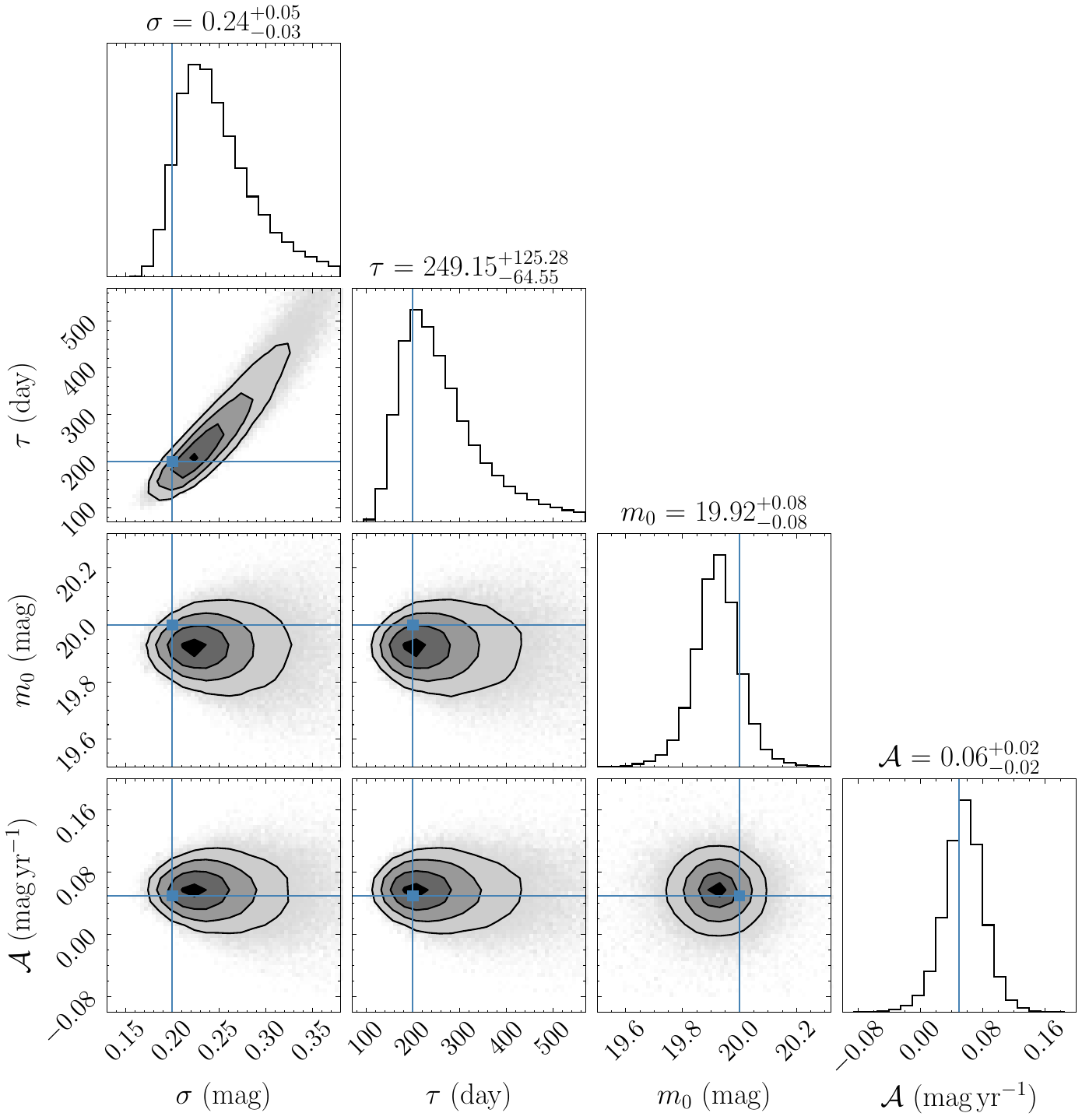}
\caption{Corner plot of the posterior of the MCMC measurement results for a simulated lensed quasar. The simulated lensed quasar has lensing signal amplitude $\mathcal{A}=0.05$\,mag\,yr$^{-1}$, quasar intrinsic variability amplitude $\sigma=0.2$\,mag, time scale $\tau=200\,$d, and quasar magnitude zero point $m_0=20$\,mag. The contours represent the 0.16, 0.5, and 0.84 percentiles, and the blue cross marks the true values of the parameters.}
\label{fig:corner}
\end{figure*}

The analytical analysis provides insights into how the detection limit varies with observational factors. To further consider the situation in realistic observations, we perform Markov chain Monte Carlo (MCMC) measurements on simulated lensed quasar light curves, and fit the amplitudes of the lensing signal with the quasar variability parameters simultaneously.

We simulate the quasar light curves as DRW processes using {\tt celerite} \citep{foreman-mackey_fast_2017}, a Python package for fast and scalable Gaussian process regression that is commonly used for modeling and fitting quasar variability. We generate light curves with variability time scale and amplitude as $\tau=200\,$d and $\sigma=0.2\,$mag, and then add a microlensing signal with amplitude $\mathcal{A}=0.05\,$mag\,yr$^{-1}$. We set the cadence of the light curve as 3 days for a time span $T=10$\,yr, mimicking the planned cadence of the LSST survey. { Additionally, we consider seasonal gaps and random missing data due to weather conditions.} To account for photometric errors, we add Gaussian white noise with a standard deviation $\sigma_0=0.05\,$mag. Figure~\ref{fig:simulated_light_curve} shows the simulated light curve with and without the microlensing signal for comparison. 

The simulated light curves are fit with the MCMC method using the {\tt emcee} package \citep{foreman-mackey_emcee_2013}, setting as free parameters the variability time scale $\tau$, amplitude $\sigma$, mean magnitude $m_0$, and microlensing amplitude $\mathcal{A}$. We adopt the likelihood function provided by the {\tt celerite} package, which implements the {\tt CARMASolver} that uses MCMC sampling to perform Bayesian inference on the probability distribution of the noise function \citep{kelly_2014_ApJ}. Notably, we find that the parameters $\sigma$ and $\tau$ show significant degeneracies that hamper the MCMC sampling, although this degeneracy is largely mitigated when adopting a new parameterization $p_1 = 2\log\,\sigma + \log\, \tau$ and $p_2 = 2\log\,\sigma - \log \,\tau$. This might be because the new parameters correspond to the low-frequency and high-frequency limits of the slopes of the logarithmic power spectral density, respectively. Choosing uniform priors for $m_0$, $A$, $p_1$, and $p_2$, equivalent to the common practice of adopting log-uniform priors for $\sigma$ and $\tau$ \citep{suberlak_improving_2021}, we run the MCMC for 5000 steps to sample the posterior, discarding the first 500 steps as the burn-in period.

\begin{figure}
\centering
\includegraphics[width=0.98\linewidth]{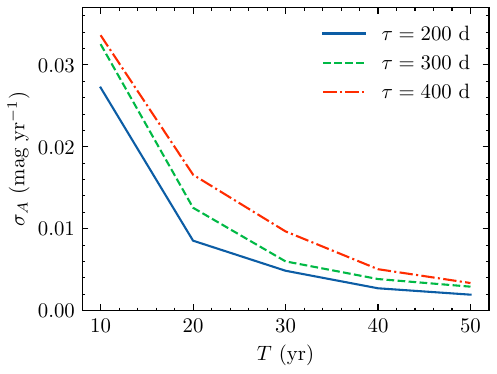}
\caption{The lensing amplitude uncertainty $\sigma_\mathcal{A}$ for different observation time span $T$ and quasar intrinsic variability time scale $\tau$. The lensing amplitude uncertainty is measured from MCMC joint fitting with the parameters of the quasars.}
\label{fig:A_func}
\end{figure}

The MCMC can recover the input parameters well. The posterior of the microlensing amplitude, $\mathcal{A}=(-0.056\pm0.027)\,$mag\,yr$^{-1}$, is consistent with the input value of $\mathcal{A}=0.050\,$mag\,yr$^{-1}$ (Figure~\ref{fig:corner}). Moreover, the obtained $\sigma_\mathcal{A}$ is consistent with our analytical prediction of $0.023\,$mag\,yr$^{-1}$ from Equation~\ref{eq:analytical}. 

To investigate how the detection limit depends on the quasar variability time scale $\tau$ and the observation time span $T$, we simulate light curves with the same procedures as above but with an array of $\tau$ and $T$. For each configuration, we simulate 10 light curves and measure their respective $\sigma_\mathcal{A}$ with MCMC; the final $\sigma_\mathcal{A}$ is the average of the 10 simulations. The distribution of $\sigma_\mathcal{A}$ obtained from the simulations (Figure~\ref{fig:A_func}) indicates that $\sigma_\mathcal{A}$ decreases with $T$ and increases with $\tau$. Moreover, the scaling relation shown in the plot is consistent with our analytical prediction of $\sigma_\mathcal{A} \propto T^{-3/2}$ and $\sigma_\mathcal{A} \propto \tau^{1/2}$. { We further explore the impact of observational cadence on lensing signal detection. As expected, our results indicate that cadence generally does not affect detectability unless it becomes comparable to $\tau$, in which case the quasar power spectrum is poorly constrained.}

To conclude: the sensitivity to long-term lensing signals from IMBHs heavily depends on the duration of the observations, with sensitivity tripling with every doubling of the observation time. Thus, it is very beneficial to combine LSST with other earlier surveys, such as the Sloan Digital Sky Survey \citep{Almeida2023ApJS} and ZTF \citep{Bellm2019PASP}, to extend the total observation time span. Meanwhile, the sensitivity is not influenced by the observational cadence once it is sufficient to constrain the quasar variability parameters. As the sensitivity decreases mildly with the quasar variability time scale, quasars with shorter variability time scales are better targets for detecting the lensing signal. High-redshift quasars may also be less effective, for cosmological time dilation makes the observed time scale appear longer, such that $\sigma_\mathcal{A}\propto (1+z)^{1/2}$. However, the mild increase may be compensated by the lower intrinsic variability time scale of quasars, in view of their lower black hole masses at earlier times.

\section{Summary}
\label{sec:summary}

We have demonstrated that the microlensing signal of IMBHs is predominantly a linear trend in the light curve, with higher order terms becoming detectable in some cases over a decade of observation. The long-term lensing signal is synchronized and exhibits the same amplitude across a range of wavelengths, from the X-rays to the ultraviolet and optical continuum. Moreover, the broad emission lines of the quasar are magnified simultaneously with the same amplitude, unless the magnification exceeds 3\,mag. 

IMBHs with compact stellar clusters may produce additional flares due to the caustics formed by the multiple lens system. The flares have time scales from months to decades and a magnification magnitude of $\sim1$\,mag. The finite source effect induces a mild wavelength dependence to the flares, on account of the color gradients of quasar accretion disks. Moreover, caustic crossing of individual broad-line clouds will result in broad emission line fluctuations, accompanied by line profile changes.

We discuss several critical observational features that can be used to distinguish microlensing events from quasar intrinsic variability and other contaminant sources:

\begin{enumerate}

\item
The microlensing long-term variability is synchronized at all wavelengths, while intrinsic variability caused by physical processes in the accretion disk with a radially stratified temperature profile exhibits time delay for different wavelengths.

\item
The microlensing long-term variability has the same amplitude in the ultraviolet, optical, and X-rays.

\item
{ Microlensing caustic-crossing events usually have a timescale of a few decades and are accompanied by broad emission-line micro-fluctuations on a timescale of hours. Additionally, the accretion disk color gradient and finite source effects cause the flares to appear earlier and fade later in redder bands than bluer bands.}

\item
The emission from the corona, but not from the dusty torus, is magnified, such that lensed quasars should violate the tight empirical correlation between mid-infrared and X-ray luminosity observed in AGNs.

\item
The magnification effects of gravitational lensing will render the quasar exceptionally bright, making them outliers in the correlation between BLR size and AGN luminosity. The apparently atypical compactness of the BLR would mimic the subset of AGNs that truly departs from the BLR size-luminosity relation for physical reasons, such as high accretion rate (e.g., \citealt{du_2016, fonseca_2020}). However, sources with genuinely high Eddington ratios can be recognized through a variety of other empirical indicators (\citealt{shen_2014}).

\item
Because the narrow-line region is too large to be magnified, the optical-ultraviolet spectrum of a lensed quasar should exhibit exceptionally weak narrow emission lines, in excess of the normal statistical trend between narrow-line equivalent width and Eddington ratio \citep{boroson_1992,shen_2014}. However, unlike AGNs of high Eddington ratio that have similarly weak narrow lines, the lensed quasars would not display other characteristics of highly accreting AGNs, such as strong broad Fe~II emission.

\item
In cases of extremely high magnification, the continuum from the accretion disk will be boosted more than the line emission from the BLR. This, too, produces a population of quasars with unusual observational properties. The differential magnification of the accretion disk relative to the BLR will result in broad emission lines with abnormally low equivalent widths, which superficially may resemble so-called weak-line quasars (e.g., \citealt{diamond_2009, plotkin_2010}). Continuum variability would not be accompanied with the normally expected corresponding response from the broad emission lines, qualitatively akin to the ``BLR holiday'' observed in some AGNs (e.g., \citealt{dehghanian_2019}).

\end{enumerate}
{
We predict that the LSST survey is capable of detecting a substantial number of IMBH microlensing events. The event number primarily depends on the variability detection limit and the cosmological mass density of IMBHs. The detection limit on the microlensing variability is sensitive to the time span of the observations, with moderate dependence on the amplitude and time scale of intrinsic quasar variability. We predict that the uncertainty of microlensing variability measurement in LSST is $\sim 0.025\,\mathrm{mag\,yr^{-1}}$ for 10 years of observation, improving to $\sim 0.01\,\mathrm{mag\,yr^{-1}}$ if we can extend the observation to 20 years.  After 10 years of monitoring, LSST could detect around 30 events above a $3\,\sigma$ variability rate threshold. This number increases by a factor of 10 if IMBHs are surrounded by stellar clusters, and it would be boosted by another order of magnitude if the survey duration extends to 20 years, provided the IMBH number density predicted by \cite{paynter_evidence_2021}.

The search for the IMBH microlensing signals, even if ending up with no positive detection, would impose stringent constraints on the cosmological mass density of IMBHs. Such constraints can be compared with previous estimation on the number density. Moreover, it will differentiate various theoretical mechanisms of IMBH formation, with strong constraints on IMBHs in the massive tail of primordial black holes and possible constraints on the prediction of cosmological hydrodynamical simulations. These results will shed light on the population of poorly known wandering IMBHs and provide critical clues for understanding the formation of black hole seeds in the early Universe and the physical processes that drive their evolution across cosmic time.

\section{Acknowledgement}
We thank the anonymous reviewer for insightful suggestions, which have significantly improved this work. We thank Jakob Robnik for advice on the matched filtering technique, and constructive comments from Jinyi Shangguan, Ziming Wang, Emma Weller, Zexuan Wu, and Ming-Yang Zhuang. LCH. was supported by the National Key R\&D Program of China (2022YFF0503401), the National Science Foundation of China (11991052, 12233001), and the China Manned Space Project (CMS-CSST-2021-A04, CMS-CSST-2021-A06). 

\appendix
\twocolumngrid
\section{Derivation of the Detection Limit}
\label{appendix:detection}  

This appendix presents the detailed derivation of the maximum likelihood estimate of the microlensing amplitude $\mathcal{A}$ and its uncertainty $\sigma_{\hat{\mathcal{A}}}$, starting from the likelihood ratio formalism. 

Given the likelihoods of the null-hypothesis in Equation~\ref{eq:log_like_H0} and the positive hypothesis in Equation~\ref{eq:log_like_H1}, the log-likelihood ratio is  

\begin{equation}  
\Lambda = \log\frac{P(s|\mathcal{H}_1)}{P(s|\mathcal{H}_0)} = \mathcal{A}\langle s, s_0\rangle - \frac{1}{2}\mathcal{A}^2\langle s_0, s_0\rangle ,
\label{eq:log_likelihood_ratio}  
\end{equation}  

\noindent
where the definition of the inner product is defined in Equation~\ref{eq:inner_product}, and $s_0(t)=t$ is the microlensing signal template. Maximizing $\Lambda$ with respect to $\mathcal{A}$ gives the maximum likelihood estimate

\begin{equation}  
\hat{\mathcal{A}} = \frac{\langle s, s_0\rangle}{\langle s_0, s_0\rangle}.  
\label{eq:mle}  
\end{equation}  

\noindent
The variance of $\hat{\mathcal{A}}$ is determined from the Fisher matrix,

\begin{equation}  
\sigma_{\hat{\mathcal{A}}}^{-2} = -\frac{\partial^2 \Lambda}{\partial \mathcal{A}^2} = \langle s_0, s_0\rangle.  
\end{equation}  

\noindent
And thus 

\begin{equation}  
\sigma_{\hat{\mathcal{A}}} = \frac{1}{\sqrt{\langle s_0, s_0\rangle}}.  
\label{eq:sigma_A}  
\end{equation}  

\noindent
For observation with duration $T$, the Fourier transform of $s_0$ is

\begin{equation}  
\begin{aligned}
\mathcal{F}\{s_0(t)\} = \int_{-T/2}^{T/2} t e^{-i2\pi ft} dt  
=\frac{i}{2} T^2 j_1(\pi fT),
\end{aligned}
\end{equation}  

\noindent
with the first-order spherical Bessel function $j_1(z) = {(\sin(z) - z\cos(z))}/{z^2}$. Substituting this result and the power spectrum in Equation~\ref{eq:power_spectrum} into $\langle s_0, s_0\rangle$ yields

\begin{equation}  
\begin{aligned}
\langle s_0, s_0\rangle &= \int_{-f_{\mathrm{max}}}^{f_{\mathrm{max}}} \frac{|\mathcal{F}\{s_0(t)\}|^2}{S_n(f)/2} df \\&= \frac{T^3}{4\sigma^2\tau} \int_0^{Tf_{\mathrm{max}}} (1 + a^2x^2){j_1^2(\pi x)} dx,  
\end{aligned}
\label{eq:inner_produect_itermediate}
\end{equation}  

\noindent
where $a = 2\pi\tau/T$, $x = fT$, and $f_{\mathrm{max}}$ is the maximum frequency used in the inference.  

The lensing signal and DRW noise are degenerate at high frequencies when $f_\mathrm{max}\gg 1/2\pi\tau$, where the power spectra of both are $\propto f^{-2}$ , as shown in Figure~\ref{fig:psd}. Therefore, the information of the lensing is primarily at low frequencies. The optimal choice should be $f_\mathrm{max}\lesssim 1/2\pi\tau$. In this regime, the term $a^2x^2$ in Equation~\ref{eq:inner_produect_itermediate} is negligible in the integration. Another constraint on $f_\mathrm{max}$ is the Nyquist sampling frequency $f_\mathrm{Nyq}=2/\Delta t$, where $\Delta t$ is the observation cadence. When $f_\mathrm{Nyq}$ is small, the inference will be largely hampered by the frequency cutoff at $f_\mathrm{Nyq}$. Therefore, to make sure that the sampling is not limited by the Nyquist frequency, we require $f_\mathrm{max}\gg 1/2\pi\tau$, meaning that $\Delta t\ll \tau$.

In the limit that $T \gg \tau$ for long-term observations, the integral in Equation~\ref{eq:inner_produect_itermediate} becomes insensitive to the exact value of $f_\mathrm{max}$ due to the rapid decay of the spherical Bessel function. Considering the limit that $\int_0^\infty j_1^2(\pi x) dx = {1}/{6}$, we obtain 

\begin{equation}  
\langle s_0, s_0\rangle  = \frac{T^3}{24\sigma^2\tau}.  
\end{equation}  
\

\noindent
Substituting into Equation~\ref{eq:sigma_A}, we find  

\begin{equation}  
\sigma_{\hat{\mathcal{A}}} = 2\sqrt{6}\sigma \tau^{1/2} T^{-3/2}.  
\label{eq:final_sigma}  
\end{equation}  
}

\end{document}